**Sequence design-based control of DNA droplets formed from phase separation of DNA nanostructures**

**Yusuke Sato,[1] Tetsuro Sakamoto,[1] Masahiro Takinoue [1]***

[1]*Department of Computer Science, Tokyo Institute of Technology, Kanagawa 226-8502, Japan*

*Correspondence to: Masahiro Takinoue (takinoue@c.titech.ac.jp)

**Abstract**

DNA has the potential to realize a controllable liquid-liquid phase separation (LLPS) system, because the design of its base sequences results in programmable interactions. Here, we have developed a novel DNA-based LLPS system which enables us to create 'DNA droplets' and to control their dynamic behaviour by designing sequences of the DNA nanostructure. We were able to change the phase separation temperature required for the formation of DNA droplets by designing the sequences. In addition, the fusion, fission, and formation of Janus-shaped droplets were controlled by sequence design and enzymatic reactions. Furthermore, modifications of proteins with sequence-designed DNAs allowed for their capture into specific droplets. Overall, our results provide a new platform for designing the phase behaviour of macromolecular structures, and paves the way for new applications of sequence-designed DNA in the creation of cell-mimicries, synthetic membraneless organelles, and artificial molecular systems.

Liquid–liquid phase separation (LLPS) of water-soluble molecules induces the formation of macromolecular droplets in water. In polymer chemistry and soft matter physics, LLPS has been studied to elucidate underlying mechanisms of phase separation[1-3] and to construct cell-like structures[4] or bioreactors[5]. With recent advances in LLPS in biological research fields, studies have found that macromolecular droplets comprised of proteins and RNAs are formed within living cells via moderate interactions among the biomolecules[6-11]. LLPS of biomolecules has garnered considerable attention as the droplets have an important role in regulating several biological functions[12]. Although several in vitro studies have reproduced droplet formation by changing the concentrations of proteins/RNAs and salts[6,9,13,14], no study has demonstrated the regulation of droplet formation depending on the biological information encoded in the biopolymer sequence. The sequence of biopolymers is an essential factor influencing droplet formation[15] because it determines intra- and inter-molecular interaction strength[16]. However, the potential of regulating LLPS by the sequence design of biopolymers remains unexplored, although it is expected that such a technology would provide new insights into not only LLPS mechanisms, but also LLPS applications.

Considerable progress is being made in modulating molecular interactions by biopolymer sequence design in the field of DNA nanotechnology, where sequence-specific hybridization of nucleic acids plays an important role. DNA nanotechnology is a promising avenue for the construction of nanostructures[17-24] and for the investigation of their phase behaviour[25-27] because the self-assembly of DNAs is precisely predictable from sequence information. The nanostructures can be further assembled into micro-sized structures. For example, a DNA hydrogel can be formed from the self-assembly of branched-DNA nanostructures[28-32], and it can serve as a substrate for cell-free protein expression[29], cytoskeleton for artificial cells[30], or carrier for drug delivery[33]. On the other hand, a few studies have reported the phase behaviour of DNA nanostructures, such as the phase separation of DNA nanostructures into DNA-poor and rich phases[25], formation of a coacervate-like DNA liquid[26], and re-entrant phase dynamics[27]. These studies focused on the effects of temperature, DNA concentration, and ionic strength. However, sequence dependency in the interaction between DNAs has not been translated into the ability to regulate phase behaviour. Sequence-design-based phase behaviour of DNA nanostructures will enable us to develop a novel LLPS system with controllability at the molecular level.

We herein report the formation and control of a 'DNA droplet' made of sequence-designed DNAs (Fig. 1). We first show that DNA nanostructures (Fig. 1a) exhibit the LLPS that results in droplet formation depending on the surrounding temperature (Fig. 1b). Next, we show how the sequence design influences the phase behaviour of DNA nanostructures. Then, we demonstrate that control of the dynamic behaviour of DNA droplets, including fusion, fission, and segregation of DNA droplets can be achieved by rational design of DNA sequences (Figs. 1c-1e). Finally, as a potential application of our DNA-based LLPS system, we demonstrate the capture and partitioning of protein-cargos in DNA droplets in a sequence-specific manner (Fig. 1f).

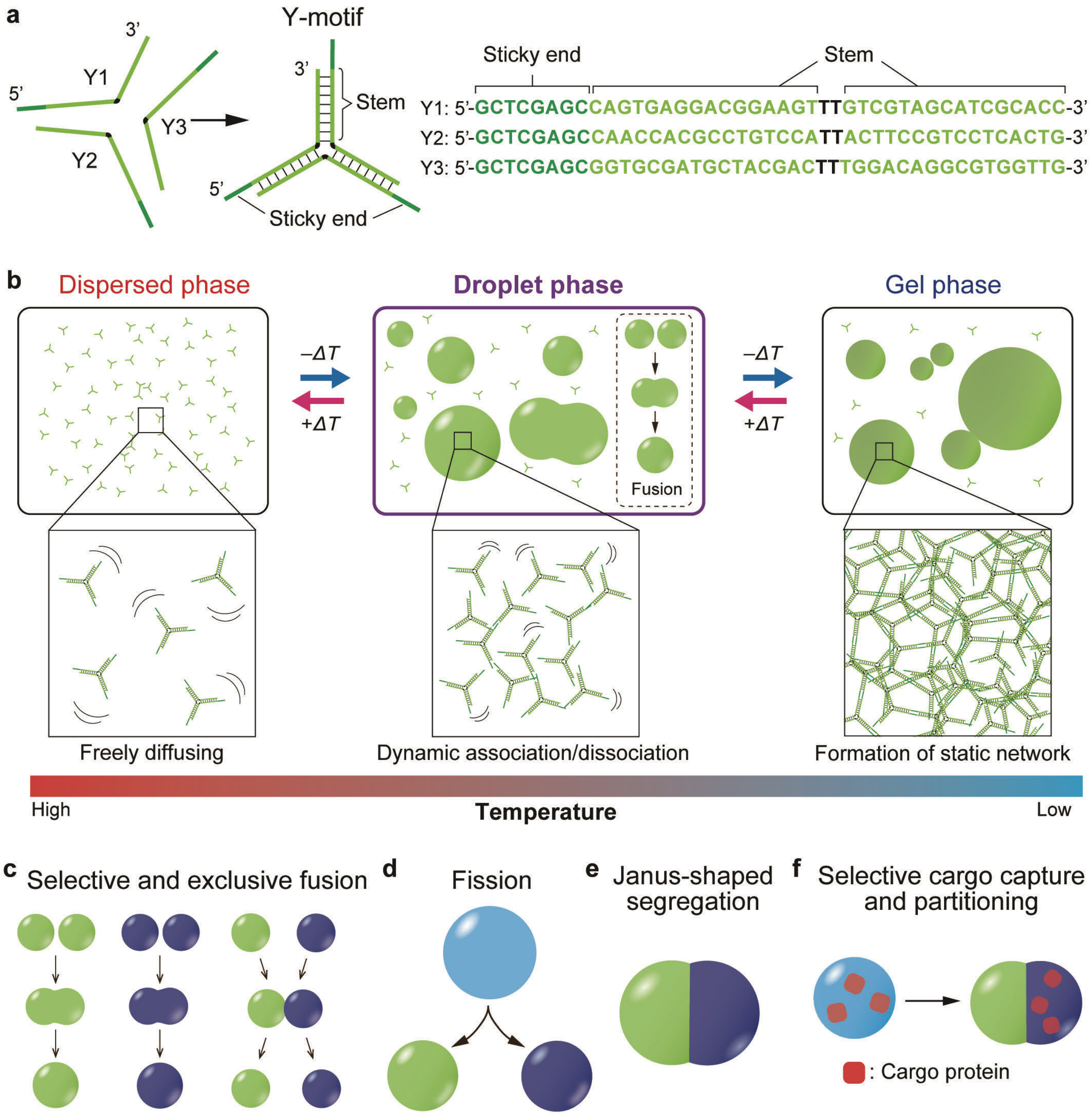

**Fig. 1. Droplet formation of DNA nanostructures based on phase separation. a**, Schematic of the Y-shaped DNA nanostructure (Y-motif), and strand sequences of the Y-motif. **b**, Temperature-dependent phase-changes of the Y-motifs. The Y-motifs are dispersed in solution at a high temperature ('dispersed phase'). Decreasing the temperature induces phase separation into a 'droplet phase' that is a liquid-liquid biphasic system composed of a Y-motif-rich liquid droplets phase in a continuous Y-motif-poor phase. By further decreasing the solution temperature, the phase changes from the droplet to a 'gel phase' that is a gel-liquid biphasic system composed of Y-motif-rich gel phase in the Y-motif-poor phase. The phase-change temperature is dominated by the sequence of DNAs. **c**, **d**, **e,** and **f**, Schematic representations for the control of DNA droplet and selective cargo capture and partitioning.

## Results

### Formation of DNA droplets via the phase separation of DNA nanostructures

We designed a Y-shaped DNA nanostructure (hereafter referred to as Y-motif) with three 'sticky ends' (SEs) (Fig. 1a). The Y-motif is composed of three single-stranded DNAs (ssDNAs): Y1, Y2, and Y3 (Fig. 1a). The 'stems' of the Y-motif have 16 base pairs that hybridized at 75 ºC in a buffer condition (20 mM Tris-HCl (pH: 8.0), 350 mM NaCl) (Supplementary Fig. 1); the SE was a palindromic 8 nucleotide (nt) sequence and the same for three branches of the Y-motif, allowing for connection between the Y-motifs.

We first observed the Y-motif solution using a confocal laser scanning microscope and a stage heater. The Y-motifs were stained with a dye molecule (SYBR Gold). At 70 °C, the Y-motifs were dispersed in a buffer solution ('dispersed phase') (Fig. 1b) and no micrometre-sized structures were observed owing to dehybridization of the SE at this temperature (Fig. 2a(I)). When the temperature was lowered to 62 °C, spherical microparticles (from several to several tens of micro-meters) were observed in the solution, although the temperature was still much higher than the melting temperature ($T_m$) of the 8 nt SE ($T_m$: 45.9 °C) (Fig. 2a(II)). Interestingly, fusion of the two particles occurred when they collided with each other (Figs. 2b and 2d, and Supplementary Movie 1); the fusion behaviour is similar to that of the droplets formed via LLPS in living cells[8]. We called the particles that exhibit such fusion 'DNA droplets', where Y-motifs spontaneously phase separate into Y-motif-rich liquid droplets in equilibrium with a Y-motif poor phase[34] ('droplet phase') (Fig. 1b). The DNA droplets grew in size via collision with one another (Supplementary Fig. 2).

To evaluate the mobility of the Y-motifs in the droplets, fluorescence recovery after photobleaching (FRAP) experiments were performed. In the FRAP experiments, the Y-motifs were labelled using FAM-modified Y2 without SE to avoid the detection of the reattaching the dye molecules to the Y-motifs. The half region of droplets was photo-bleached and the changes in fluorescence intensity were measured in the bleached regions. The fluorescence intensity in the droplets was immediately recovered after the bleaching (Figs. 2f and 2h), which suggests that the Y-motifs can diffuse inside droplets (do not form static networks) and the dynamic dissociation and association of the Y-motifs result in the formation of the droplet.

When the temperature was further decreased to 25 °C (Fig. 2a(III)), the microparticles did not fuse (Figs. 2c and 2e). In addition, in the FRAP experiments, the fluorescence intensity did not immediately recover at around this temperature (Figs. 2g and 2i). Thus, the Y-motifs formed static networks and the DNA droplets became Y-motif-rich hydrogels ('gel phase') (Fig. 1b). A previous study using a microrheological technique reported that DNA nanostructures exhibited a gradual transition into a hydrogel with decreasing temperatures[27]. Our results suggest that DNA nanostructures transition through a droplet phase before gelation.

We further found that this phase behaviour between the dispersed, droplet, and gel phases was reversible. The hydrogels changed into droplets when temperatures increased to 62 °C (Fig. 2a (IV)), and the Y-motifs dispersed at 70 °C (Fig. 2a(V)). These temperature-dependent phase-changes were confirmed with further continuous temperature changes (Supplementary Fig. 3).

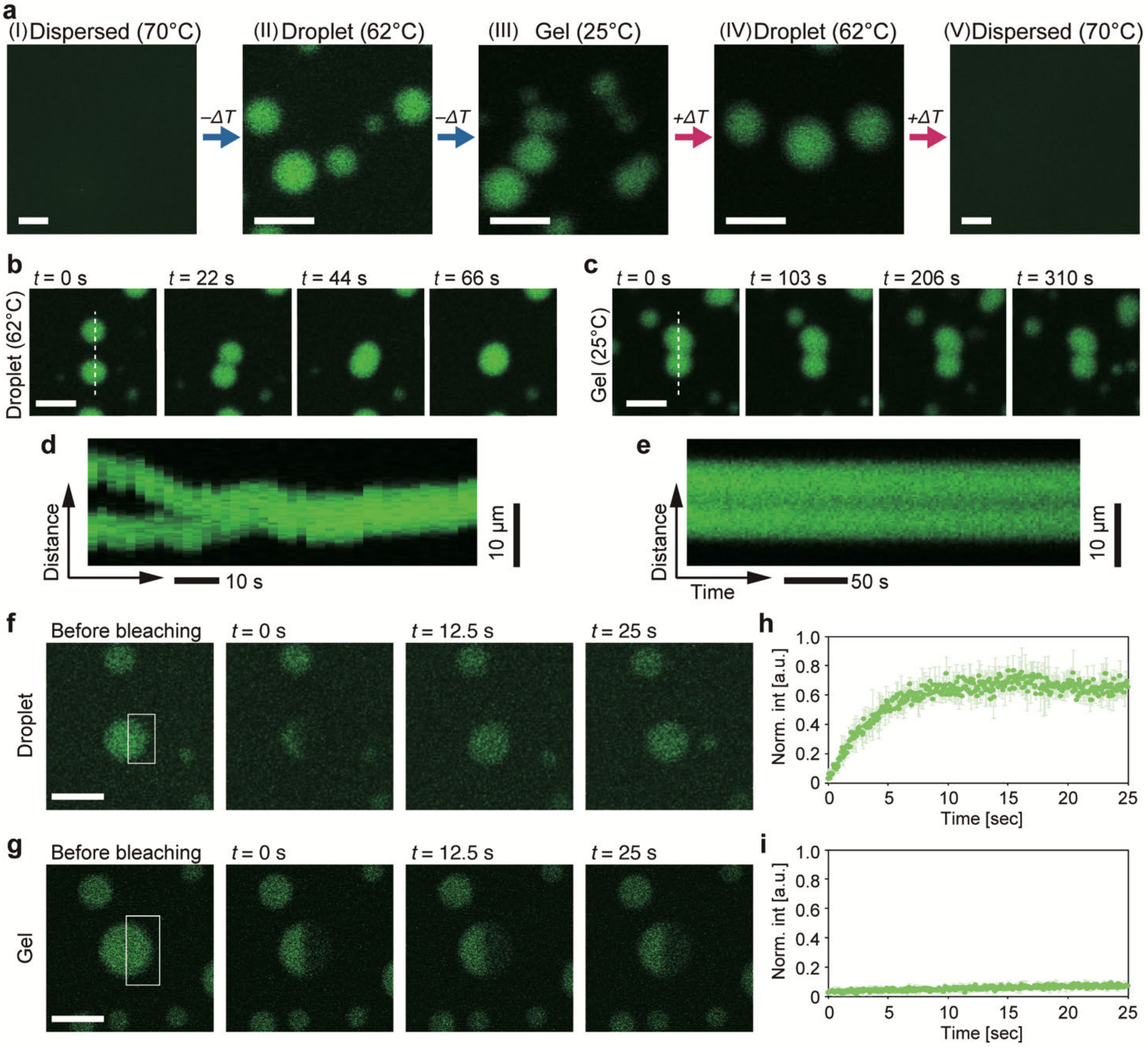

**Fig. 2. Microscopic observation of DNA droplets and hydrogels.** **a**, Microscopic images for reversible phase-changes in Y-motifs with 8 nucleotide long sticky ends, between the dispersed, droplet, and gel phases, in response to the surrounding temperature. **b** and **c**, Sequential images for DNA droplets (**b**) and hydrogels (**c**) at 62 and 25 °C, respectively. **d** and **e**, Kymographs along the dashed line shown in the image at $t = 0$ s in **b** and **c**. **f** and **g**, Image sequences of fluorescence recovery after photobleaching for DNA droplets (**f**) and hydrogels (**c**), visualized at 60 and 30 °C, respectively. The bleached region is indicated by a white box. **h** and **i**, Time series of fluorescence intensity in DNA droplets (**h**) and hydrogels (**i**). Error bars represent the mean ± standard deviation (S.D.). The number of measurements ($n$) = 3. Scale bars: 100 µm in (I) and (V) in **a**; 10 µm in other images.

**Effects of the sequence and the number of branches on phase separations and changes**

We next investigated the effects of the SE sequences on the phase-change temperatures between the dispersed and droplet phases ($T_d$), and between the droplet and gel phases ($T_g$). We defined $T_d$ as the temperature at which droplet formation was observed (Supplementary Fig. 4), and $T_g$ as the temperature at which less recovery of fluorescence in the FRAP experiment was detected (Supplementary Fig. 5). $T_d$ and $T_g$ were identified by decreasing the temperature from 85 ºC at a rate of –1 ºC/min. The phase diagram (Fig. 3a) was experimentally built for the Y-motifs with 4, 6, 8, 10, and 12 nt SEs (Fig. 3b and Supplementary Table 1). The phase diagram showed that larger $-\Delta H$ (longer sticky end lengths) tended to result in an increase of both of $T_d$ (solid line in Fig. 3a) and $T_g$ (dashed line in Fig. 3a); $T_d$ were 46.3, 48.3, 63.7, 61.3, and 70.3 °C and $T_g$ were 12.3, 18.7, 35.7, 41.3, and 53.3 °C, respectively, when the SEs were 4-12 nt long. Note that droplet formation was observed for the Y-motifs whose SEs that were 4-12 nt long (Supplementary Fig. 6), but not with those that were 2 nt SE ($\Delta H$: –9.6 kcal/mol) (Supplementary Fig. 7).

We also investigated the effects of the number of branches in the motifs on the $T_d$ and $T_g$. We additionally designed two motifs that had four or six branches with 8 nts long SEs (Supplementary Table 2). The experimental results showed that the $T_d$ and $T_g$ increased with the number of branches in the motifs: 67.7 and 55.3 °C for the motifs having four branches and 72.3 and 61.3 °C for the motifs having six branches (Fig. 3c, Supplementary Figs. 8-10).

The fact that the DNA droplets were formed at higher temperatures than the $T_m$ of the SEs suggests that unstable hybridization between the SEs plays a key role in the formation of the DNA droplets. Such weak interactions of the SEs would allow for the dynamic association/dissociation of the motifs (Figs. 2f and 2h), which leads to the fluidic behaviour of DNA droplets, e.g. fusion of the droplets. On the other hand, the increase of the number of branches in a motif resulted in the increase of $T_d$ (Fig. 3c). Thus, both the stability of hybridization in the SEs and the number of branches in a motif would determine $T_d$ (see Supplementary Figs. 11-16 and Supplementary Discussion)[35, 36].

We considered that $T_g$ was attributed to the stability of SEs and entanglement of motifs. Previous studies have reported that the gelation process in the branched-DNA nanostructures occurs at around the $T_m$ of SEs[32]. On the other hand, our results show that $T_g$ for the Y-motif with 4 nt SE was higher than the $T_m$ of the 4 nt SE (Fig. 3a). DNA nanostructures can be concentrated by the phase separation, and concentration of the DNA changes $T_m$. Hence, we experimentally estimated the concentration of the DNA nanostructures in the gel phase (Supplementary Figs. 17 and 18). Under the estimated concentration, the $T_m$ of each SE exhibited a higher value than the $T_g$ (Supplementary Fig. 12) and the stability of the hybridization of SEs would be enough to form static networks (Supplementary Fig. 19). The increase of $T_g$ with the number of branches in the motif (Fig. 3c) could be explained by entanglement of the motifs. We determined the $T_g$ based on the FRAP experiments, which showed mobility of the motifs. The mobility can be decreased by the hybridization of the SEs between the motifs, and an increase in the number of branches in a motif induces the structural entanglements that also decrease the mobility of the motifs. The $T_g$ in motifs having four or six branches was higher than their $T_m$ (Fig. 3c), and the $T_g$ in the six SE motifs was higher than that in four. Therefore, the effects of the entanglement on $T_g$ would become more dominant than the stability of the hybridization of the SEs.

**a**

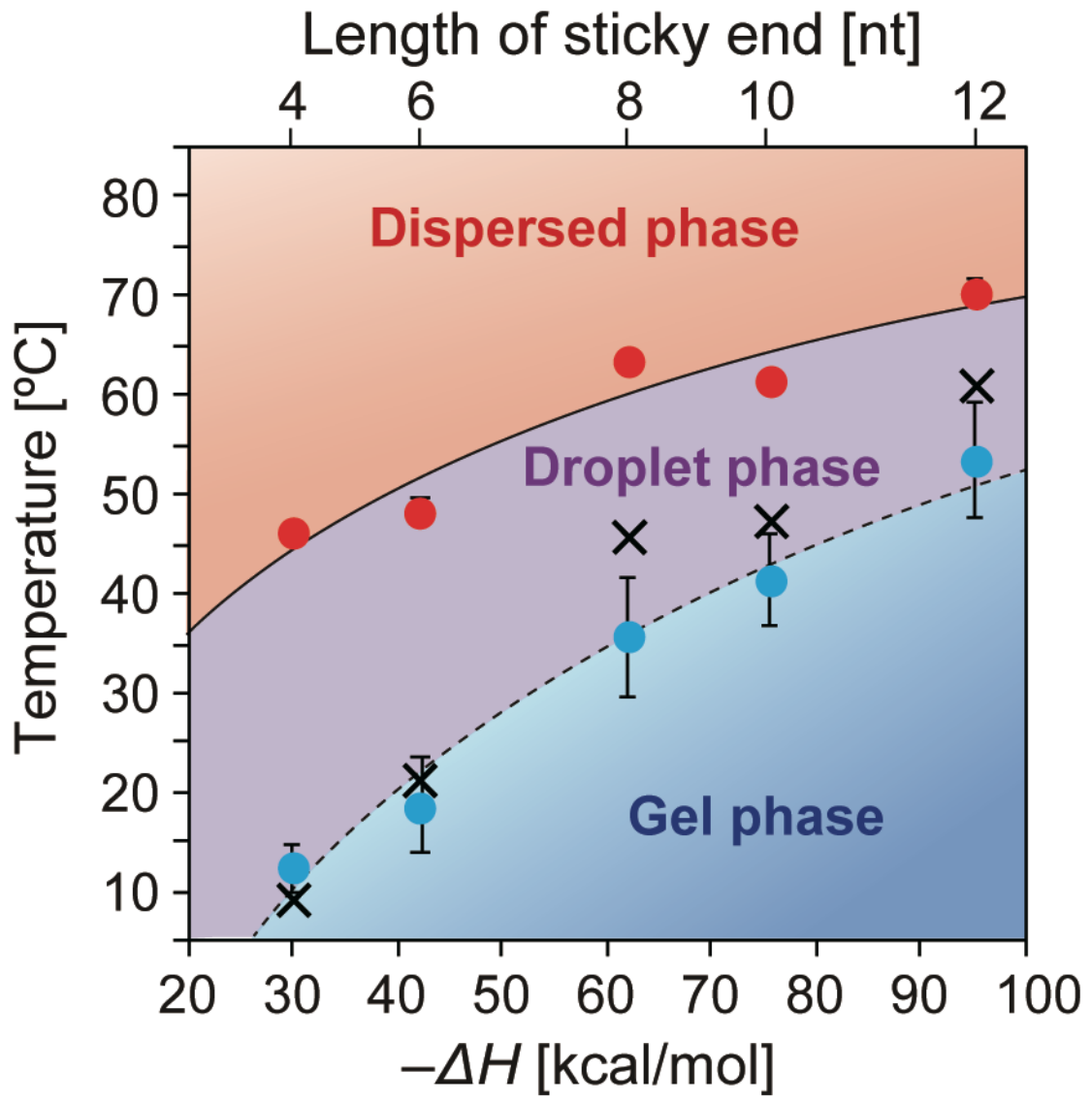

**b**

| Sticky end (5'→3') | Length [nt] | –ΔH [kcal/mol] | $T_m$ [°C] |
|---|---|---|---|
| GCTAGCGCTAGC | 12 | 95.2 | 61.1 |
| GACTCGAGTC | 10 | 75.6 | 47.6 |
| GCTCGAGC | 8 | 62.0 | 45.9 |
| GCTAGC | 6 | 42.2 | 21.5 |
| GCGC | 4 | 30.0 | 9.3 |

**c**

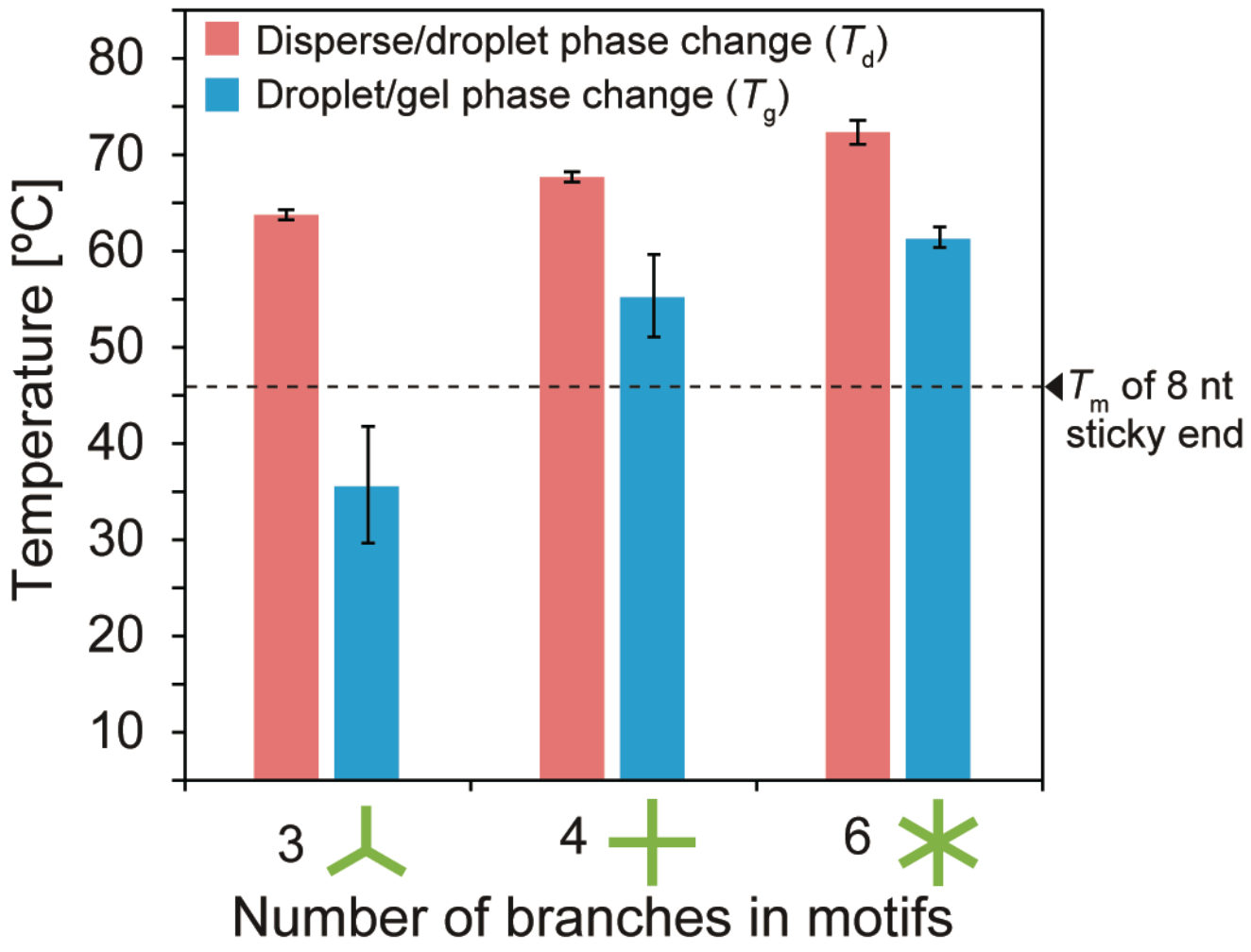

**Fig. 3. Sticky end-dependent phase behaviour of DNA nanostructures. a**, Phase diagram of Y-motifs with different lengths of sticky end. Red and blue dots represent mean values of the phase-change temperatures between the dispersed and droplet phases ($T_d$), and between the droplet and gel phases ($T_g$), respectively. Cross-marks in the diagram show melting temperatures ($T_m$) of each SE calculated in the condition of 15 μM Y-motifs and 350 mM $Na^+$ ions (corresponds to the experimental condition). Error bars indicate standard deviation (mean ± S.D., $n$ = 3). Solid and dashed lines are guides. **b**, List of sticky end sequences, lengths, enthalpy changes ($-\Delta H$), and $T_m$. **c**, $T_d$ and $T_g$ of a motif with different numbers of branches. Error bars indicate standard deviation (mean ± S.D., $n$ = 3). Length of the sticky end was fixed at 8 nucleotides (nt) ($-\Delta H$: 62.0 kcal/mol, $T_m$: 45.9 ºC). Horizontal dashed line indicates the $T_m$ of 8 nt SE.

**Orthogonality in DNA droplets**

We expected that the rational design of the motif sequence allowed for the control of the droplet-droplet interactions. We designed an 'orthogonal Y-motif' with 8 nt SE ($^{orth}$Y-motif) whose sequences were not complementary to the Y-motif with 8 nt SE, but had almost similar thermodynamic parameters ($\Delta H$ and $T_m$) in their stem and SEs (Supplementary Tables 3 and 4). The orthogonality of the SE sequences between the Y- and $^{orth}$Y-motifs enables us to acknowledge the sequence-specific fusion of DNA droplets (Fig. 4a). The Y- and $^{orth}$Y-motifs were labelled with FAM (green) and Alexa 405 (blue), respectively. Then all strands for both of the motifs were mixed in a solution, and the motifs were annealed on the stage heater from 85 °C at a rate of –1 °C/min to 60 °C.

As we expected, the Y- and $^{orth}$Y-motifs were individually phase separated into droplets. Fusion events were observed between the droplets composed of the same motifs but not between the droplets formed by different motifs (Fig. 4b and Supplementary Movie 2). To confirm the selective and exclusive fusion was derived from the sequence orthogonality, we further designed a six-junction motif (S-motif) with two types of SEs corresponding to the SEs of the Y- and $^{orth}$Y-motif (Fig. 4c and Supplementary Table 5). The S-motifs play a role for the cross-bridging between the Y- and $^{orth}$Y-motifs and can eliminate the orthogonality between the two types of motifs, which results in the formation of DNA droplets composed of Y-, $^{orth}$Y- and S-motifs (Fig. 4d). The Cy5-labelled S-motif, FAM-labelled Y-motif, and Alexa405-labelled $^{orth}$Y-motif were mixed in a 1:3:3 molar ratio and visualized at 65 °C. In the experiment, the fluorescence of Cy5, FAM, and Alexa405 were observed in one droplet, suggesting that the Y- and $^{orth}$Y-motifs were mixed via the S-motifs in the droplets (Fig. 4e). The elimination of the orthogonality was also confirmed in a two-dimensional histogram of FAM/Alexa405 fluorescence intensities (Supplementary Fig. 20). In addition, the droplets composed of Y-, $^{orth}$Y- and S-motifs exhibited fusion (Fig. 4f) similar to the droplets of a sole motif composition. Thus, selective fusion of the DNA droplets was achieved in two motifs that were orthogonal and the addition of the motif for cross-bridging eliminated the selectivity.

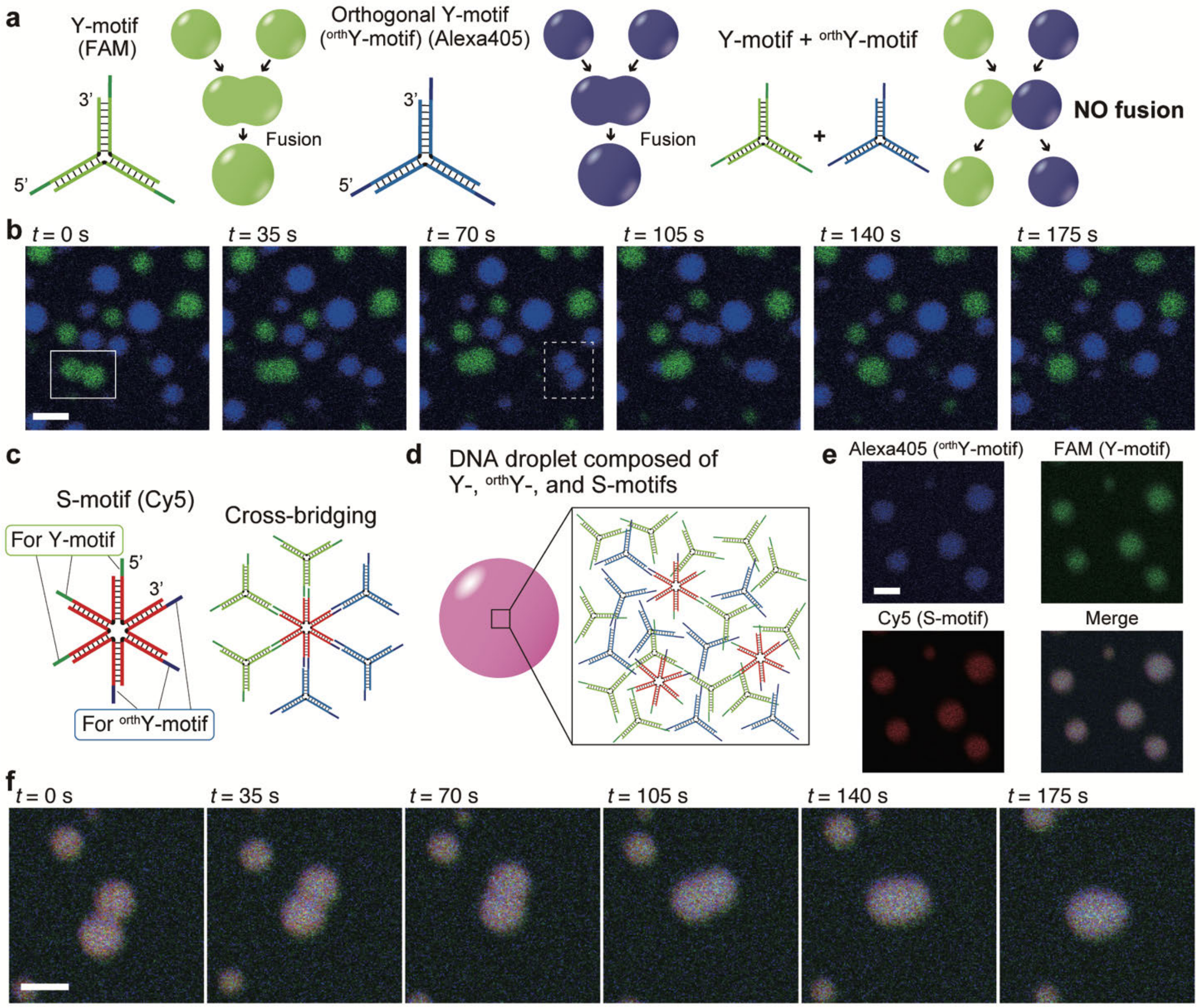

**Fig. 4. Sequence-dependent fusion control in DNA droplets.** **a**, Schematic illustrations of the selective fusion in the Y- and orthogonal Y-motifs ($^{orth}$Y-motif). **b**, Time series of microscopic images representing fusion events for the droplets composed of the Y-motifs (box in solid line at $t$ = 0 s) and $^{orth}$Y-motifs (box in dashed line at $t$ = 70 s). Blue and green channels indicate $^{orth}$Y- (Alexa405) and Y-motifs (FAM), respectively. **c,** Six-junction motif (S-motif) having sticky ends for the Y- and $^{orth}$Y-motifs. **d**, Schematic of DNA droplets composed of the Y-, $^{orth}$Y-, and S-motifs. **e** and **f**, Microscopic images and time series of the fusion for DNA droplets composed of the Y-, $^{orth}$Y-, and S-motifs. Scale bars: 10 μm.

**Fission and segregation of DNA droplets**

To further demonstrate the controllability of the DNA droplets, we designed a mechanism for their fission (Figs. 5a and 5b). DNA-RNA chimera strands were introduced into the S-motif, termed 'chimerized-S-motif (CS-motif)' that also has SEs for both the Y- and $^{orth}$Y-motifs (Fig. 5a and Supplementary Table 5). Two RNA regions were located at around the centre of the CS-motif and ribonuclease A (RNase A) decomposes the RNA parts. The enzymatic reaction results in splitting CS-motif into two three-branched portions with affinity to the Y-motif ('Y-portion') and the $^{orth}$Y-motif ('$^{orth}$Y -portion') (Fig. 5a, Supplementary Fig. 21). After splitting, the cross-bridge ability of the CS-motif disappears, and the Y- and $^{orth}$Y-motifs cannot be mixed owing to their intrinsic orthogonality. As a result, the DNA droplets exhibit micro-phase separation in a spinodal decomposition manner, and finally fission into multiple droplets composed of Y-motif/Y-portion or $^{orth}$Y-motif/$^{orth}$Y-portion (Fig. 5b).

The Y-, $^{orth}$Y-, and CS-motifs formed droplets, representing the mixing of the Y- and $^{orth}$Y-motifs via CS-motifs. (Fig. 5c). Real-time monitoring of the RNase-induced split of the CS-motif successfully visualized the process of the fission of DNA droplets (Fig. 5d and Supplementary Movie 3). After the addition of the RNase A to the sample solution during the observation, the Y- and $^{orth}$Y-motifs gradually separated inside droplets ($t$ = 30 s in Fig. 5d and Supplementary Fig. 22), and finally exhibited complete fission into two or three droplets ($t$ = 150 s in Fig. 5d and Supplementary Fig. 23). Line profiles for each fluorophore showed the colocalization of Y-motifs and Y-portions, and $^{orth}$Y-motifs and $^{orth}$Y-portions (Fig. 5e), indicating the successful split of the CS-motifs as designed. The addition of only the buffer solution did not induce such fission behaviour (Supplementary Fig. 24).

By using S-motifs (non-cleavable by RNase A) in addition to CS-motifs (cleavable by RNase A), we demonstrated the formation of complex-shaped droplets. When the ratio of the CS-motif was large but a few S-motifs existed, addition of RNase A did not induce fission but caused the segregation of the droplets owing to the immiscibility and the partial cross-bridging by a few S-motifs (Fig. 5f). Experimental results showed that in the 90 % ratio of the CS-motifs, the DNA droplets segregated and finally formed Janus-shaped droplets that have two portions for the Y- and $^{\text{orth}}$Y-motifs in the droplets (Fig. 5g and Supplementary Figs. 25 and 26). Fluorescence of Cy3 ($^{\text{orth}}$Y-portions) and Cy5 (Y-portions) were colocalized to Alexa405 ($^{\text{orth}}$Y-motif) and FAM (Y-motif), respectively. Furthermore, the 50 % ratio of the CS-motifs resulted in the patchy-like pattern formation in the DNA droplets (Supplementary Figs. 27 and 28). These results showed the existence ratio of the S- and CS-motifs determined the shape of the DNA droplets after the addition of the RNase A, and indicated that the S-motifs played a role for not only the cross-bridging but also for a ‘surfactant’ between the Y- and $^{\text{orth}}$Y-motifs.

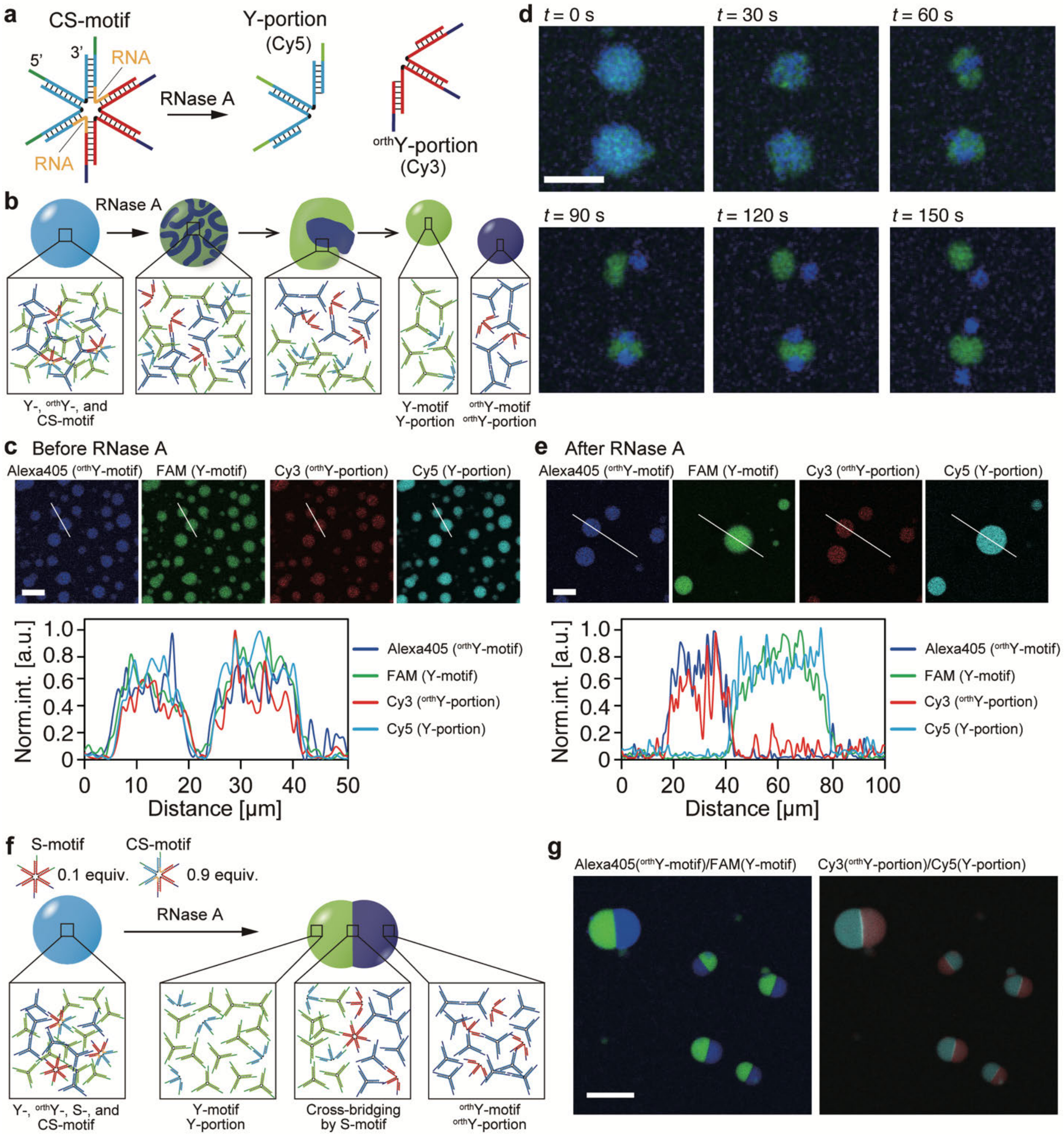

**Fig. 5. Fission and segregation of DNA droplets based on sequence design and enzymatic reaction. a**, Six-junction motif containing DNA-RNA chimera strands (CS-motif). RNA sequences are introduced at around the centre part of the CS-motif. Enzymatic reactions of ribonuclease A (RNase A) results in the split of the CS-motif into two portions, termed Y- and $^{orth}$Y-portion. Y- and $^{orth}$Y-portion were labelled with Cy5 and Cy3, respectively **b**, Fission process of DNA droplets and the motifs that comprise the droplets. **c**, Microscopic images of DNA droplets before the addition of RNase A for Alexa405 ($^{orth}$Y-motif), FAM (Y-motif), Cy3 ($^{orth}$Y-portion), and Cy5 (Y-portion). The bottom graph shows the normalized intensity profile of each fluorophore on the line shown in the above images. **d**, Time series images of fission of DNA droplets. Channels for Alexa405 (blue) and FAM (green) were merged. **e**, Microscopic images of DNA droplets and line profile for each fluorophore after the addition of RNase A. **f,** Schematic of Janus-shaped DNA droplet. At the middle of the Janus DNA droplet, a few S-motifs cross-bridge the immiscible motif groups (Y-motifs/Y-portions and $^{orth}$Y-motifs/$^{orth}$Y-portions). **g,** Microscopic images for Alexa405/FAM (blue/green) and Cy3/Cy5 (red/cyan) channels in the Janus-shaped DNA droplets. Scale bars: 20 μm in **d**; 30 μm in other images.

**Selective cargo capture and partitioning in DNA droplets**

The combination of our DNA-based LLPS system and other molecules would provide an application basis and increase the functionality of the DNA droplets. We focused on the capture of cargo into the DNA droplets via DNA-modification. DNA-streptavidin conjugates that have SEs for Y- or $^{orth}$Y-motifs were prepared (Fig. 6a) and the selective capture of streptavidin in the DNA droplets was visualized (Fig. 6b).

DNA-modified streptavidin, labelled with DyLight 549 fluorophores, was added during the visualization of DNA droplets composed of Y- or $^{orth}$Y-motifs. The streptavidin was successfully localized in specific DNA droplets, depending on the modified DNA sequence (Figs. 6c and 6d). In addition, the accumulation processes of the streptavidin into DNA droplets were also visualized (Supplementary Fig. 29).

Moreover, by adopting the technique to prepare the Janus-shaped DNA droplets (Fig. 5f), partitioning of the cargos in DNA droplets was achieved (Fig. 6e-6j). Streptavidin, modified with SE sequences, were homogeneously distributed in DNA droplets in which the Y- and $^{orth}$Y-motifs were mixed by the S- and CS-motifs (Fig. 6f and 6i). After the addition of RNase A, DNA droplets segregated into the Janus shape, and streptavidin were partitioned into the complementary sides depending on the modified sequence (Fig. 6g and 6j). As we demonstrated here, modification of targets with SE sequences, which encode interaction information, allowed us to capture them in DNA droplets and to control their partitioning in a sequence-dependent manner. These results suggest that DNA droplets have a potential to be carriers of designated molecules and to provide reaction environments for biochemical processes as artificial membraneless organelles.

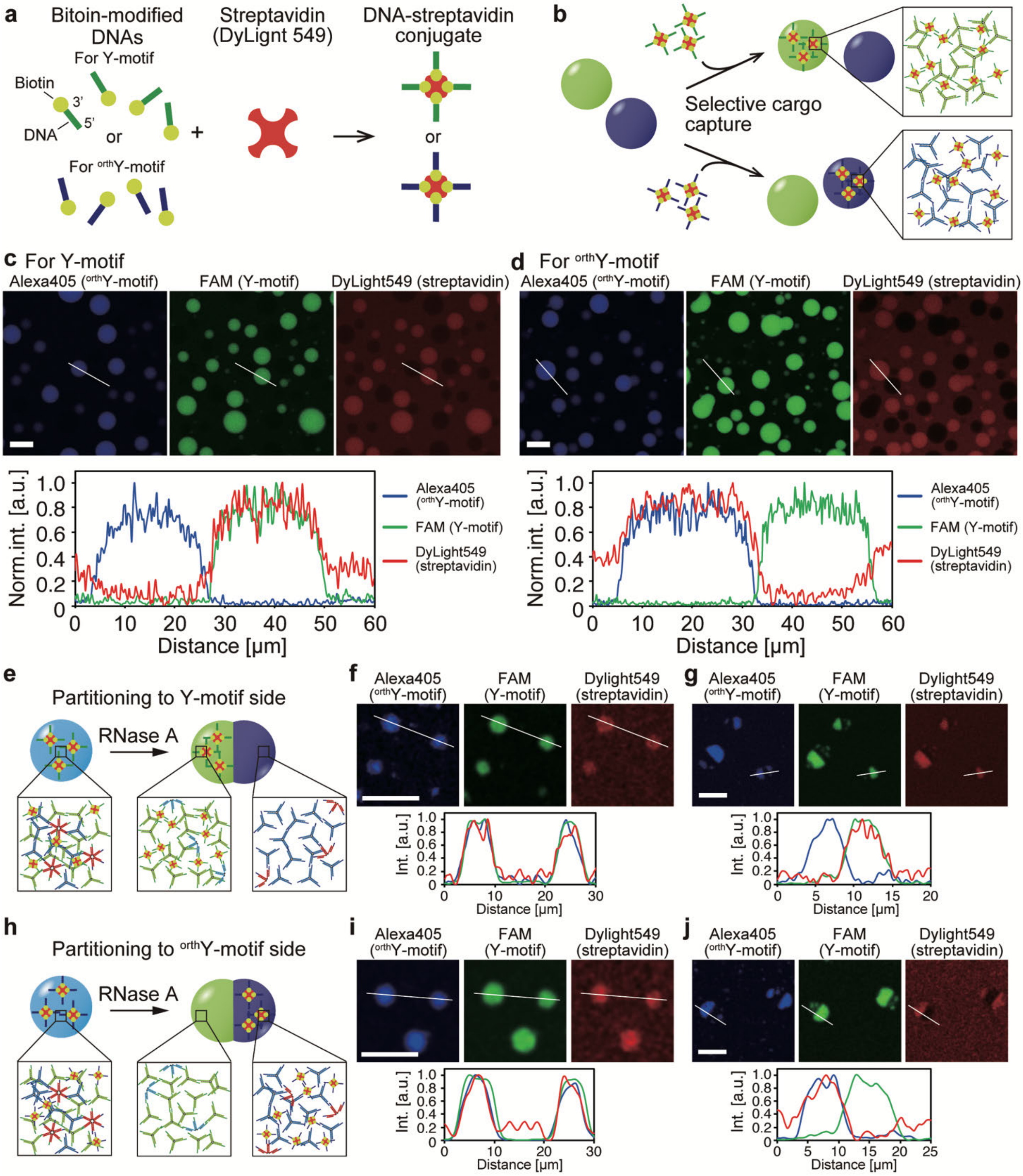

**Fig. 6. Protein capture and partitioning in DNA droplets.** **a**, Streptavidin modified with biotinylated-DNAs which have sticky end (SE) sequence for Y- and orthogonal Y-motifs ($^{orth}$Y-motifs). **b,** The DNA-modified streptavidin can be captured in DNA droplets in the sequence-dependent manner. **c** and **d,** Microscopic images and line profiles representing localization of streptavidin modified with sticky end for Y- (**c**) or for $^{orth}$Y-motifs (**d**). The bottom graphs show normalized intensity profile of each fluorophore on the line shown in the above images. **e** and **h,** Schematics of partitioning DNA-modified streptavidin to the portion composed of Y- (**e**) or $^{orth}$Y-motifs (**h**). **f** and **g,** Microscopic images of DNA droplets and streptavidin modified with sticky end for Y-motifs before (**f**) and after the addition of RNase A (**g**). The bottom graphs show normalized intensity profile of each fluorophore on the line shown in the above images. **i** and **j,** Microscopic images of DNA droplets and streptavidin modified with sticky end for $^{orth}$Y-motifs before (**i**) and after the addition of RNase A (**j**). The bottom graphs show normalized intensity profile of each fluorophore on the line shown in the above images. Scale bars: 30 μm in **c** and **d**; 20 μm in other images.

## Discussion

In this paper, we reported the sequence-dependent phase behaviour of the DNA nanostructures and the control of DNA droplets according to the information encoded in the SE sequences. The sequence-dependent phase-changes were reversible between the dispersed, droplet, and gel phases in response to the surrounding temperature (Fig. 2a), which is similar to the gas-liquid-solid phase transition of matter. We expect that these findings will be applicable to other nanostructures such as DNA origami[22], DNA-modified proteins[37], or nanoparticles covered with DNA strands[38-40]. In other words, the phase behaviour shown in the present study could be expanded to matters that can be modified with DNAs, which may enable us to design the phase and create droplets for any material.

A remarkable point in our DNA droplet system is the controllability of the interaction and dynamic behaviour of the droplets (Figs. 3-5). As living systems are well organized-dynamic structures whose behaviour is regulated by the information encoded in biopolymers[41], our DNA-based LLPS system would provide a new basis for the development of artificial cell engineering. Additionally, our demonstration for the capture of proteins in DNA droplets with sequence specificity (Fig. 6) can also make significant contributions to synthetic biology to generate a biochemical reaction environment composed of designed molecules.

From the viewpoint of material science, the successful creation of the Janus-shaped DNA droplets (Fig. 5g) would give us a fundamental technique to construct water-soluble complex-shaped droplets in an aqueous phase, which is intrinsically different from the conventional Janus-droplets[42] based on the water/oil immiscibility. Furthermore, the Janus DNA droplets were created owing to a surfactant-like role of DNA nanostructures for the cross-bridge of the two types of Y-motif. It suggests that multi-affinity surfactants would be achieved by the modification of DNAs to designated molecules.

Finally, we envision that their combination with biochemical circuits that can be programmed to produce and decompose specific DNA strands[43-45], will lead to the design and control of autonomous dynamic behaviour of DNA droplets as if living organisms. Ultimately, such autonomous behaviour of macromolecular structures could serve for the development of molecular robotic systems[46, 47], which become comparable with living cellular systems in the future. We believe that this study is an important step toward the development of a novel LLPS system whose dynamical behaviour can be controlled by information encoded in component molecules.

## Methods

### Oligonucleotide preparation

Sequences for the DNA and DNA-RNA chimera strands are provided in the Supplementary Tables 1-3, and 5. The oligonucleotides were purchased from Eurofins Genomics (Tokyo, Japan) and were of OPC grade purification, except for the fluorophore-labelled and chimerized strands, that were purified to an HPLC grade. The oligonucleotides were dissolved in ultrapure water (18 MΩ cm in resistance) at 100 µM concentrations and stored at –20 ºC until use.

### Preparation of the observation chamber

Glasses and coverslips, measuring 30 × 40 mm and 18 × 18 mm, respectively, with a thickness of 0.17 mm, were purchased from Matsunami Glass (Kishiwada, Japan). The glasses (30 × 40) were treated with oxygen plasma using a plasma cleaning machine (PIB-20, Vacuum Device, Japan); and were then soaked in 5 % (w/v) BSA, dissolved in 20 mM Tris-HCl (pH 8.0), for over 30 min, to prevent non-specific interactions of the DNA on the glass surface. After the BSA coating, the glasses were washed with distilled water and dried under an air-flow. The BSA-coated glasses and the cover slips (18 × 18 mm) were assembled using double-sided tape. The sample solutions were immersed into the slit between the coated glass and the cover slips and the edges were sealed with manicure paste. To avoid evaporation of the sample solution during the observation period, the chamber was further covered with a mineral oil (Nacalai tesque, Kyoto, Japan) using a bank made of silicone sheet.

### Visualization

Samples in the observation chamber were visualized using a confocal laser scanning microscope

(FV1000, Olympus, Tokyo, Japan) and a stage heater (10021-PE120 system, Japan high tech (Linkam), Hukuoka, Japan). SYBR Gold and FAM were visualized at excitation wavelengths of 473 nm. Alexa405, Cy3, and Cy5 were visualized at 405, 559, and 635 nm, respectively.

**Repetitive phase-change of Y-motifs**

DNA strands for the Y-motifs with 8 nucleotide long sticky ends were mixed in a test tube with the buffer, 5 μM for each strand. SYBR Gold was also added at 1 × concentration to the test tube to stain the DNA. The test tube was heated at 85 ºC for 3 min; then gradually cooled to 25 °C at a rate of –1 °C/min, using a thermal cycler (Mastercycler nexus X2, Eppendorf, Hamburg, Germany). The annealed sample was first visualized at 25 °C and then the temperature was changed to 70 °C at a rate of 20 °C/min, and then fluorescence images were obtained. The temperature was then decreased to 62 °C and droplet formation via the phase separation was sequentially visualized at a scanning velocity of 2.2 s per frame. Subsequently, the temperature was further changed to 25, 62, 70, 62, and 25 °C, in this order, at a rate of 20 °C/min. The samples were visualized at each temperature. A detailed-procedure of the repetitive phase-change and representative microscopic images for each temperature are shown in the Supplementary Methods and the Supplementary Fig. 3.

**Measurement of the phase-change temperatures and FRAP experiments**

Each motif was labelled with dye-modified DNA without sticky ends, that was mixed at a 10 % molar ratio in the solution. (e.g., 5 μM for Y-1 and Y-3, 4.5 μM for Y-2, and 0.5 μM for Y2_0_FAM). The sample solution was poured in the chamber on the stage heater and incubated at 85 °C for 3 min. Subsequently, the temperature was decreased to a designated temperature at a rate of –1 °C/min. To determine the temperature at which the DNA droplets were formed, the sample was visualized at the 1 °C interval. To evaluate the phase-change from the droplet phase to the gel phase, the fluidity of the motifs in the particles formed by the phase separation was analysed using FRAP experiments. A rectangular-shaped region in the DNA droplets or hydrogels was photobleached at the respective wavelength for each fluorophore. Sequential images were obtained at a scanning velocity of 65 ms per frame. The average fluorescence intensity values for the fluorescence in the bleached region were measured at designated temperatures.

**Reaction with ribonuclease A**

Ribonuclease A (RNase A) (Nippon Gene, Toyama, Japan) was added to sample solutions to a final concentration of 20 μg/ml. The samples were visualized at 62 or 65 °C, where motifs can exhibit droplet-like behaviour. Detailed methods are provided in the Supplementary Methods.

**Capture and partitioning of DNA-modified streptavidin**

DyLight 549-labelled streptavidin was purchased from Vector Laboratories (CA, USA). Biotin molecules were modified to the 3` end in DNA with the sticky end sequence. The streptavidin and the biotinylated DNA were mixed in a sample solution for a final concentration of 20 μg/ml and 1.5 μM.

**Acknowledgements**

We would like to thank to Mr. Hiroki Watanabe for his kind support in the data analysis. This research was supported by JSPS KAKENHI to M.T. (No. JP17H01813, JP18K19834) and Y.S. (No. JP18J00720), Research Encouragement Grants from The Asahi Glass Foundation to M.T., and Support for Tokyotech Advanced Researchers to M.T. Y.S. is a JSPS Research Fellow (SPD).

**Author contributions**

M.T. provided the original concept. Y.S. and M.T. planned the experiments. Y.S. performed all experiments. T.S. aided in data analysis. Y.S. and M.T. wrote the manuscript.

## Competing financial interests

The authors declare that they have no competing financial interests.

# Supplementary Information

**Sequence design-based control of DNA droplets formed from phase separation of DNA nanostructures**

**Yusuke Sato,[1] Tetsuro Sakamoto,[1] Masahiro Takinoue [1]***

[1]*Department of Computing Science, Tokyo Institute of Technology, Kanagawa 226-8502, Japan*

*Correspondence to: Masahiro Takinoue (takinoue@c.titech.ac.jp)

## Contents

## Supplementary Methods

### Oligonucleotide sequences

The sequences of the oligonucleotide strands were designed using a web software, NUPACK[1] (http://www.nupack.org/). Sequences are shown in Supplementary Tables 1-3, and 5. To ensure the efficient formation of motifs, the flexibility of the stem was introduced by inserting two spacer bases (TT or rUrU) at the centre of the junction. Each strand was named N-*i*_*l*_M, where 'N' represents the name of the motif for which the strand is used, '*i*' a strand identification number, '*l*' the length of the sticky end (in nucleotides), and 'M' a modified fluorescent molecule. In the Tables, the sticky ends and modified fluorescent molecules are shown in black, the spacer bases in grey, and complementary parts in the stems in the same colour.

**Supplementary Table 1.** Y-motif sequences

| Name | Sequence (5` – 3`) |
|---|---|
| Y-1_12 | GCTAGCGCTAGCCAGTGAGGACGGAAGTTTGTCGTAGCATCGCACC |
| Y-2_12 | GCTAGCGCTAGCCAACCACGCCTGTCCATTACTTCCGTCCTCACTG |
| Y-3_12 | GCTAGCGCTAGCGGTGCGATGCTACGACTTTGGACAGGCGTGGTTG |
| Y-1_10 | GACTCGAGTCCAGTGAGGACGGAAGTTTGTCGTAGCATCGCACC |
| Y-2_10 | GACTCGAGTCCAACCACGCCTGTCCATTACTTCCGTCCTCACTG |
| Y-3_10 | GACTCGAGTCGGTGCGATGCTACGACTTTGGACAGGCGTGGTTG |
| Y-1_8 | GCTCGAGCCAGTGAGGACGGAAGTTTGTCGTAGCATCGCACC |
| Y-2_8 | GCTCGAGCCAACCACGCCTGTCCATTACTTCCGTCCTCACTG |
| Y-3_8 | GCTCGAGCGGTGCGATGCTACGACTTTGGACAGGCGTGGTTG |
| Y-1_6 | GCTAGCCAGTGAGGACGGAAGTTTGTCGTAGCATCGCACC |
| Y-2_6 | GCTAGCCAACCACGCCTGTCCATTACTTCCGTCCTCACTG |
| Y-3_6 | GCTAGCGGTGCGATGCTACGACTTTGGACAGGCGTGGTTG |
| Y-1_4 | GCGCCAGTGAGGACGGAAGTTTGTCGTAGCATCGCACC |
| Y-2_4 | GCGCCAACCACGCCTGTCCATTACTTCCGTCCTCACTG |
| Y-3_4 | GCGCGGTGCGATGCTACGACTTTGGACAGGCGTGGTTG |
| Y-1_2 | GCCAGTGAGGACGGAAGTTTGTCGTAGCATCGCACC |
| Y-2_2 | GCCAACCACGCCTGTCCATTACTTCCGTCCTCACTG |
| Y-3_2 | GCGGTGCGATGCTACGACTTTGGACAGGCGTGGTTG |
| Y-2_0_FAM | [6-FAM]-CAACCACGCCTGTCCATTACTTCCGTCCTCACTG |

**Supplementary Table 2.** Sequences of motifs with four and six branches with 8 nucleotide long sticky ends

| Name | Sequence (5` – 3`) |
|---|---|
| Four-1_8 | GCTCGAGCGCTGGACTAACGGAACGGTTAGTCAGGTATGCCAGCAC |
| Four-2_8 | GCTCGAGCGTGCTGGCATACCTGACTTTCGCAAATTTACAGCGCCG |
| Four-3_8 | GCTCGAGCCGGCGCTGTAAATTTGCGTTCATCACTTGGGACCATGG |
| Four-4_8 | GCTCGAGCCCATGGTCCCAAGTGATGTTCCGTTCCGTTAGTCCAGC |
| Four-2_0_Cy3 | [Cy3]-GTGCTGGCATACCTGACTTTCGCAAATTTACAGCGCCG |
| Six-1_8 | GCTCGAGCGCTGGACTAACGGAACGGTTAGTCAGGTATGCCAGCAC |
| Six-2_8 | GCTCGAGCCTCAGAGAGGTGACAGCATTCCGTTCCGTTAGTCCAGC |
| Six-3_8 | GCTCGAGCCCATGGTCCCAAGTGATGTTTGCTGTCACCTCTCTGAG |
| Six-4_8 | GCTCGAGCCGGCGCTGTAAATTTGCGTTCATCACTTGGGACCATGG |
| Six-5_8 | GCTCGAGCCAGACGTCACTCTCCAACTTCGCAAATTTACAGCGCCG |
| Six-6_8 | GCTCGAGCGTGCTGGCATACCTGACTTTGTTGGAGAGTGACGTCTG |
| Six-5_0_Cy5 | [Cy5]-CAGACGTCACTCTCCAACTTCGCAAATTTACAGCGCCG |

**Supplementary Table 3.** Orthogonal Y-motif ($^{orth}$Y-motif) sequences

| Name | Sequence (5` – 3`) |
|---|---|
| $^{orth}$Y-1_8 | **CTCGCGAG**AAAGGAACTCTCCGCGTTGACAAAGCCGACACGT |
| $^{orth}$Y-2 _8 | **CTCGCGAG**GCCTCTGTGTCGCATCTTCGCGGAGAGTTCCTTT |
| $^{orth}$Y-3 _8 | **CTCGCGAG**ACGTGTCGGCTTTGTCTTGATGCGACACAGAGGC |
| $^{orth}$Y-2 _0 _Alexa405 | [Alexa405]-GCCTCTGTGTCGCATCTTCGCGGAGAGTTCCTTT |

**Supplementary Table 4.** $T_m$ and enthalpy changes ($\Delta H$) of the sequences for the Y- and $^{orth}$Y-motifs. These parameters were obtained using DINAMelt[2] with 350 mM $Na^+$ and 5 μM DNA (for the stems) and 15 μM DNA (for the sticky ends)

| Sequence pair | $T_m$ [°C] | $-\Delta H$ [kcal/mol] |
|---|---|---|
| CAGTGAGGACGGAAGT vs ACTTCCGTCCTCACTG | 63.5 | 122.3 |
| GTCGTAGCATCGCACC vs GGTGCGATGCTACGAC | 66.1 | 129.4 |
| CAACCACGCCTGTCCA vs TGGACAGGCGTGGTTG | 67.1 | 129.6 |
| AAAGGAACTCTCCGCG vs CGCGGAGAGTTCCTTT | 64.8 | 124.7 |
| GACAAAGCCGACACGT vs ACGTGTCGGCTTTGTC | 65.7 | 127.2 |
| GCCTCTGTGTCGCATC vs GATGCGACACAGAGGC | 66.0 | 127.7 |
| GCTCGAGC vs GCTCGAGC | 45.9 | 62.0 |
| **CTCGCGAG** vs **CTCGCGAG** | 45.5 | 62.8 |

**Supplementary Table 5.** Sequences of six-junction motifs (S-motif) and the chimerized-S-motif (CS-motif)

| Name | Sequence |
|---|---|
| S-1_8 | **CTCGCGAG**GCTGGACTAACGGAACGGTTAGTCAGGTATGCCAGCAC |
| S-2_8 | **CTCGCGAG**CTCAGAGAGGTGACAGCATTCCGTTCCGTTAGTCCAGC |
| S-3_8 | **CTCGCGAG**CCATGGTCCCAAGTGATGTTTGCTGTCACCTCTCTGAG |
| S-4_8 | GCTCGAGCCGGCGCTGTAAATTTGCGTTCATCACTTGGGACCATGG |
| S-5_8 | GCTCGAGCCAGACGTCACTCTCCAACTTCGCAAATTTACAGCGCCG |
| S-6_8 | GCTCGAGCGTGCTGGCATACCTGACTTTGTTGGAGAGTGACGTCTG |
| S-2_0_Cy3 | [Cy3]-CTCAGAGAGGTGACAGCATTCCGTTCCGTTAGTCCAGC |
| S-5_0_Cy5 | [Cy5]-CAGACGTCACTCTCCAACTTCGCAAATTTACAGCGCCG |
| CS-1_8 | **CTCGCGAG**GCTGGACTAACGGArArCrGrGrUrUrArGrUrCAGGTATGCCAGCAC |
| CS-4_8 | GCTCGAGCCGGCGCTGTAAATTrUrGrCrGrUrUrCrArUrCACTTGGGACCATGG |

The letter "r" precedes a ribonucleotide.

**Measurement of $T_m$ for the Y-motif**

DNA strands for the Y-motifs without sticky ends were mixed in a test tube at 5 µM of each strand with a buffer, consisting of 20 mM Tris-HCl (pH 8.0) and 350 mM NaCl. SYBR Green I (Takara Bio, Kusatsu, Japan) was added at 1× concentration to the test tube to detect the formation of the double helix. The test tubes were set on a real-time PCR detection system (CFX Connect, Bio-Rad, Hercules, CA, USA). The sample was heated at 85 ºC for 3 min; then cooled to 5 °C at a rate of –1 °C/min. Subsequently, the temperature was increased from 5 °C to 85 °C at the same rate. The fluorescence intensity was measured at 1 °C intervals during the decrease and increase of temperature (Supplementary Fig. 1).

**Concentration estimation**

The concentrations of the Y-motifs in the droplets and gel phases were estimated by measuring the fluorescence intensity of the supernatant, after centrifugation, that precipitates the DNA droplets and hydrogels. First, the calibration curve for the intensity-based concentration values was prepared using Y-motifs without sticky ends (composed of Y-1_0, Y-2_0, Y-3_0 and 10 %(mol) Y-2_0_FAM). The Y-motif without a sticky end was annealed at 5 μM in a 20 μl volume in the PCR test tube with the buffer, using the thermal cycler from 85 °C to 4 °C, at a rate of –1 °C/min. The annealed Y-motifs without sticky ends were then diluted to 100, 10, and 1 nM for a 20 μl volume using the buffer, and the fluorescence intensity was measured using a spectrometer (FP-6500, JASCO, Hachioji, Japan). Next, the Y-motifs with 4, 6, 8, 10, and 12 nucleotide long sticky ends were also annealed by the same procedure. After annealing, to precipitate the DNA droplets and hydrogels at the bottom of the test tubes, the PCR tube was placed in a larger test tube as a support during centrifugation (Supplementary Fig. 18 a). Then, the tubes were centrifuged at 2000 × *g* at 4 °C for 20 min. The precipitation of the phase separated Y-motifs were confirmed using a gel imager (Gel Doc EZ system, Bio-Rad, Hercules, CA, USA) (Supplementary Fig. 18 b). The PCR tubes containing the precipitate were set on the thermal cycler. For measuring the concentrations in the droplet and gel phases, the temperatures were set at 1 °C lower than the $T_d$ for each length of the sticky end or 4 °C, respectively. The test tubes were incubated for 10 min at each temperature, and the supernatant was collected using a gel loading tip (BM Equipment, Tokyo, Japan). Ten micro-litres of supernatant was diluted to 200 μl using the buffer, and the fluorescence intensity was measured. The concentration of Y-motifs was estimated using the calibration curve.

**Observations of orthogonality in the DNA droplets and its elimination**

For the orthogonality experiments, DNA strands were mixed in a test tubes with the buffer at the concentrations shown in Table 6. This mixture was poured in the observation chamber and set on the stage heater. Then, the temperature was kept at 85 °C for 3 min, and decreased at a rate of −1 °C/min. The fusion events of the DNA droplets were visualized at 60 °C using the confocal microscope at a scanning velocity of 7 s per frame.

To confirm the elimination of the orthogonality, the Y-, $^{orth}$Y-, and S-motif were mixed in a test tube with the buffer at the concentrations shown in Table 7. The mixture was then visualized using the same method as described above.

Supplementary Table 6: Strand concentration analysis to confirm orthogonality

| DNA strand name | Final concentration |
|---|---|
| Y-1_8 | 5.0 μM |
| Y-2_8 | 4.5 μM |
| Y-3_8 | 5.0 μM |
| Y-2_0_FAM | 0.5 μM |
| $^{orth}$Y-1_8 | 5.0 μM |
| $^{orth}$Y-2_8 | 4.5 μM |
| $^{orth}$Y-3_8 | 5.0 μM |
| $^{orth}$Y-2_0_Alexa405 | 0.5 μM |

Supplementary Table 7: Strand concentration for the elimination of orthogonality

| DNA strand name | Final concentration |
|---|---|
| Y-1_8 | 5.000 µM |
| Y-2_8 | 4.500 µM |
| Y-3_8 | 5.000 µM |
| Y-2_0_FAM | 0.500 µM |
| $^{orth}$Y-1_8 | 5.000 µM |
| $^{orth}$Y-2_8 | 4.500 µM |
| $^{orth}$Y-3_8 | 5.000 µM |
| $^{orth}$Y-2_0_Alexa405 | 0.500 µM |
| S-1_8 | 1.650 µM |
| S-2_8 | 1.650 µM |
| S-3_8 | 1.650 µM |
| S-4_8 | 1.650 µM |
| S-5_8 | 1.485 µM |
| S-6_8 | 1.650 µM |
| S-5_0_Cy5 | 0.165 µM |

**Confirmation of the cleavage of DNA-RNA chimera strands by ribonuclease A**

Cleavage of the DNA-RNA chimera strands was confirmed with denaturing urea polyacrylamide gel electrophoresis. A denaturing 10 % polyacrylamide gel containing 8 M urea was prepared by mixing acrylamide/Bis (29:1) solution (Nacarai tesque, Kyoto, Japan), urea (FUJIFILM Wako Pure Chemical, Osaka, Japan), tris-borate EDTA buffer (Nippon Gene, Toyama, Japan), ammonium peroxodisulfate (FUJIFILM Wako Pure Chemical), and N,N,N',N'-tetramethylethylenediamine (FUJIFILM Wako Pure Chemical). The DNAs, DNA-RNA chimera strands, CS-motifs, and S-motifs (all of them were annealed) were mixed with 20 µg/ml of ribonuclease A (RNase A) (Nippon Gene, Toyama, Japan) at 5 µM concentrations in test tubes. They were incubated at 37 °C for 15 min; then, the sample solutions were mixed with a loading buffer, composed of 0.05 % (w/v) bromophenol blue, 1 mM EDTA (Nippon Gene), and ~100 % (v/v) deionized formamide (FUJIFILM Wako Pure Chemical), and further incubated at 95 °C for 10 min. The electrophoresis was performed at 200 V for 30 min at 25 °C. The gel was stained with SYBR Gold and imaged using the gel imager.

**Fission of DNA droplets**

To achieve fission of the DNA droplets, part of the DNA strands constituting the S-motif were replaced with DNA-RNA chimera strands, considering the motif geometry for the split into the Y- and $^{orth}$Y- portions. DNA and DNA-RNA chimera strands were mixed at the concentrations given in Supplementary Table 8. The samples were then annealed from 85 °C to 25 °C at a rate of –1 °C/min using the thermal cycler. Twenty microliters of the annealed samples were put on BSA-coated glass, on which a silicone sheet with an 8 mm punch hole in diameter was placed. Mineral oil was poured on the sample in the pore to avoid evaporation of the sample during the observations. The glass was put on the stage heater and incubated at 60 °C for 30 min so that the sample solution was fully in the droplet phase. RNase A was dissolved in the buffer and was added to the sample solution at the final concentration of 20 µg/ml during the observations at 65 °C. As the control experiment, the buffer solution was added to the sample instead of the RNase A solution.

Supplementary Table 8: Strand concentrations for droplet fission

| DNA strand name | Final concentration |
|---|---|
| Y-1_8 | 5.000 μM |
| Y-2_8 | 4.500 μM |
| Y-3_8 | 5.000 μM |
| Y-2_0_FAM | 0.500 μM |
| $^{orth}$Y-1_8 | 5.000 μM |
| $^{orth}$Y -2_8 | 4.500 μM |
| $^{orth}$Y-3_8 | 5.000 μM |
| $^{orth}$Y-2_0_Alexa405 | 0.500 μM |
| CS-1_8 | 1.650 μM |
| S-2_8 | 1.485 μM |
| S-3_8 | 1.650 μM |
| CS-4_8 | 1.650 μM |
| S-5_8 | 1.485 μM |
| S-6_8 | 1.650 μM |
| S-2_0_Cy3 | 0.165 μM |
| S-5_0_Cy5 | 0.165 μM |

**Segregation of DNA droplets**

The Janus-shape and patchy-like patterns in the DNA droplets, were created by adjusting the concentrations of DNA-RNA chimera strands. For the Janus-shape and patchy-like pattern, 90 and 50 mol % of the DNA strands were replaced with the chimera strands, respectively (Supplementary Table 9 and 10). Sample solutions in the test tubes were annealed from 85 °C to 25 °C at a rate of –1 °C/min using the thermal cycler; then, 1 µl of the RNase A solution was added to the solutions for the 20 µg/ml final concentration. These mixtures were poured into the observation chamber on the stage heater and the temperature was increased to 62 °C, and visualized using the confocal microscope.

Supplementary Table 9: Strand concentrations for the Janus-shaped DNA droplets

| DNA strand name | Final concentration |
|---|---|
| Y-1_8 | 5.000 µM |
| Y-2_8 | 4.500 µM |
| Y-3_8 | 5.000 µM |
| Y-2_0_FAM | 0.500 µM |
| $^{\mathrm{orth}}$Y-1_8 | 5.000 µM |
| $^{\mathrm{orth}}$Y -2_8 | 4.500 µM |
| $^{\mathrm{orth}}$Y-3_8 | 5.000 µM |
| $^{\mathrm{orth}}$Y-2_0_Alexa405 | 0.500 µM |
| CS-1_8 | 0.165 µM |
| S-1_8 | 1.485 µM |
| S-2_8 | 1.485 µM |
| S-3_8 | 1.650 µM |
| CS-4_8 | 0.165 µM |
| S-4_8 | 1.485 µM |
| S-5_8 | 1.485 µM |
| S-6_8 | 1.650 µM |
| S-2_0_Cy3 | 0.165 µM |
| S-5_0_Cy5 | 0.165 µM |

Supplementary Table 10: Strand concentration for the patchy-like pattern in the DNA droplets

| DNA strand name | Final concentration |
|---|---|
| Y-1_8 | 5.000 µM |
| Y-2_8 | 4.500 µM |
| Y-3_8 | 5.000 µM |
| Y-2_0_FAM | 0.500 µM |
| $^{orth}$Y-1_8 | 5.000 µM |
| $^{orth}$Y -2_8 | 4.500 µM |
| $^{orth}$Y-3_8 | 5.000 µM |
| $^{orth}$Y-2_0_Alexa405 | 0.500 µM |
| CS-1_8 | 0.825 µM |
| S-1_8 | 0.825 µM |
| S-2_8 | 1.485 µM |
| S-3_8 | 1.650 µM |
| CS-4_8 | 0.825 µM |
| S-4_8 | 0.825 µM |
| S-5_8 | 1.485 µM |
| S-6_8 | 1.650 µM |
| S-2_0_Cy3 | 0.165 µM |
| S-5_0_Cy5 | 0.165 µM |

## Supplementary Discussion

The phase diagram (Fig. 3a in the main text) showed that the phase-change temperature between the dispersed and droplet phase ($T_d$) for each sticky end (SE) was higher than the $T_m$ of the SEs. The $T_m$ is generally defined as the temperature at which half of the DNA strands in a solution formed double helixes when two complementary DNA strands were present. On the other hand, our motifs have multiple SEs (e.g., Y-motif has three SEs). Thus, the $T_m$ for the SEs should not be considered, but instead, the $T_m$ for the motifs ($T_{m\text{-Motif}}$) at which half of the motifs in a solution are connected, should be considered.

In the branched motifs, the hybridization of one of multiple SEs would be sufficient for the connection between the motifs. Based on this idea, we considered that the $T_{m\text{-Motif}}$ can be described as the temperature at which at least one of the multiple SEs can hybridize at a 50 % ratio. It is calculated by the following equation:

$$1 - \left(1 - X(T)\right)^k = 0.5,$$

where $X(T)$ is the hybridization probability of SEs at the temperature $T$; and $k$ is the number of branches in a motif. The $T_{m\text{-Motif}}$ is calculated by obtaining an $X(T)$ that satisfied this equation.

The $X(T)$ in the motifs with three, four, and six branches ($k$ = 3, 4, and 6) were 21, 16, and 11 %, respectively. It means that the temperature at which the hybridization probability of SEs in each number of branches exhibit those percentage corresponds to the $T_{m\text{-Motif}}$ (Supplementary Fig. 11) For example, in the case of the motif having three branches (Y-motif) with 4 nucleotide (nt) long SE, the 18.8 ºC resulted in the 21 % of hybridization probability (obtained by using NUPACK[1]), which represents the $T_{m\text{-Motif}}$ of the Y-motif with 4 nt SEs. Similarly, the $T_{m\text{-Motif}}$ of the Y-motifs with 6, 8, 10, and 12 nt SEs were 29.2, 53.1, 59.3, and 68.0 °C, respectively (Supplementary Fig. 12). Those $T_{m\text{-Motif}}$ values were closer to the $T_d$ than that of the $T_m$ for the SEs in the bulk solution. Furthermore, the $T_{m\text{-Motif}}$ increased with the number of branches; 53.1 °C for three branches, 55.8 °C for four branches, and 59.6 °C for six branches (Supplementary Fig. 13); this trend is consistent with the experimental results (Fig. 3c). However, there were still gaps between the $T_d$ and $T_{m\text{-Motif}}$, in particular, for the shorter SEs, which suggests that another parameter can influence the phase-change temperature.

The phase diagram also showed that the difference between the $T_d$ and $T_g$ (the phase-change temperature between the droplet and gel phase) becomes larger with decreasing SE length. As the hybridization stability of the SEs would play a key role in determining the phase-change temperature of both the $T_d$ and $T_g$, the width of the melting range ($T_w$) provides an indication for the phase-change temperature.

In addition to $T_m$, $T_w$ is also an important parameter for the characterization of melting behaviour. $T_w$ can be defined as the difference between the temperatures at the end and the start of DNA melting (Supplementary Fig. 14). DNA strands transfer from ssDNA to dsDNA states in the range of $T_w$ and *vice-versa*. At the start of the melting, SEs can exhibit weak interactions, and at the end of melting, SEs form stable double helixes. As the self-assembly of motifs into droplets and gel structures results from weak interactions and stable network-formations, we supposed that temperature differences between $T_d$ and $T_g$ ($\Delta T_{\text{droplet-gel}}$) related to $T_w$.

We calculated $T_w$ for each of the SEs from the melting curve using NUPACK[1] (Supplementary Fig. 15). The $\Delta T_{\text{droplet-gel}}$ and $T_w$ were shown in Supplementary Fig. 16. $\Delta T_{\text{droplet-gel}}$ and $T_w$ exhibited closer values and the differences between $\Delta T_{\text{droplet-gel}}$ and $T_w$ are smaller in the shorter SEs. Although the entanglement of motifs can influence $T_g$, as we discussed in the main text, this graph suggests that $T_w$ may be one of the parameters that determines the phase-change temperature.

Taken together, the entanglements of motifs, multiple SEs in motifs that provide new parameters of $T_{\text{m-Motif}}$, and $T_w$ of SEs could be possible factors influencing the phase-change temperature and their combinatorial effects may determine $T_d$ and $T_g$. Model-based numerical simulations could be a feasible approach to further understand what determines $T_d$ and $T_g$, that will be utilized in our future reports.

## Supplementary Figs. 1-29

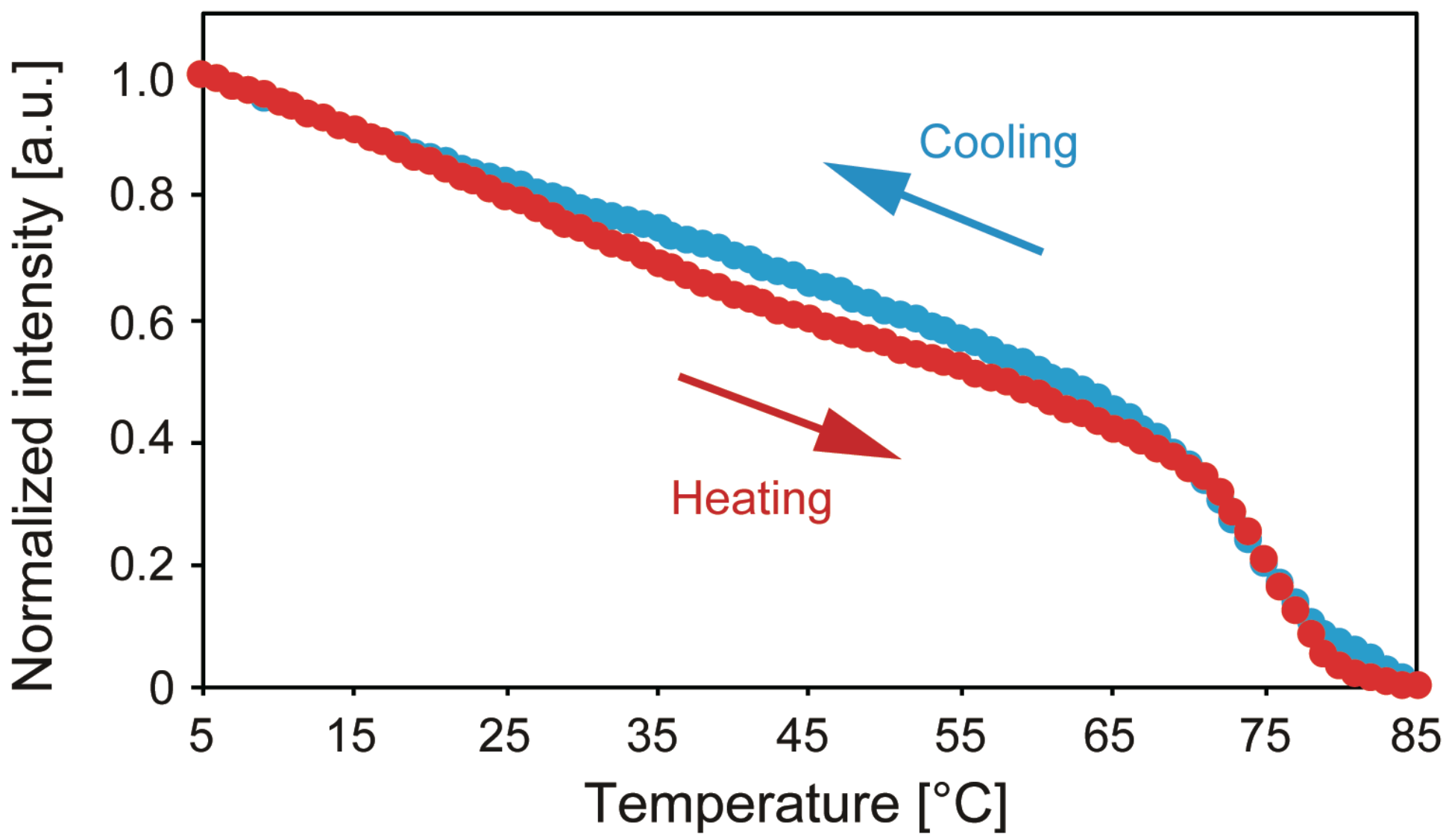

**Supplementary Fig. 1**. Normalized fluorescence intensity during the cooling and heating processes for Y-motifs without sticky ends that cannot connect to one another.

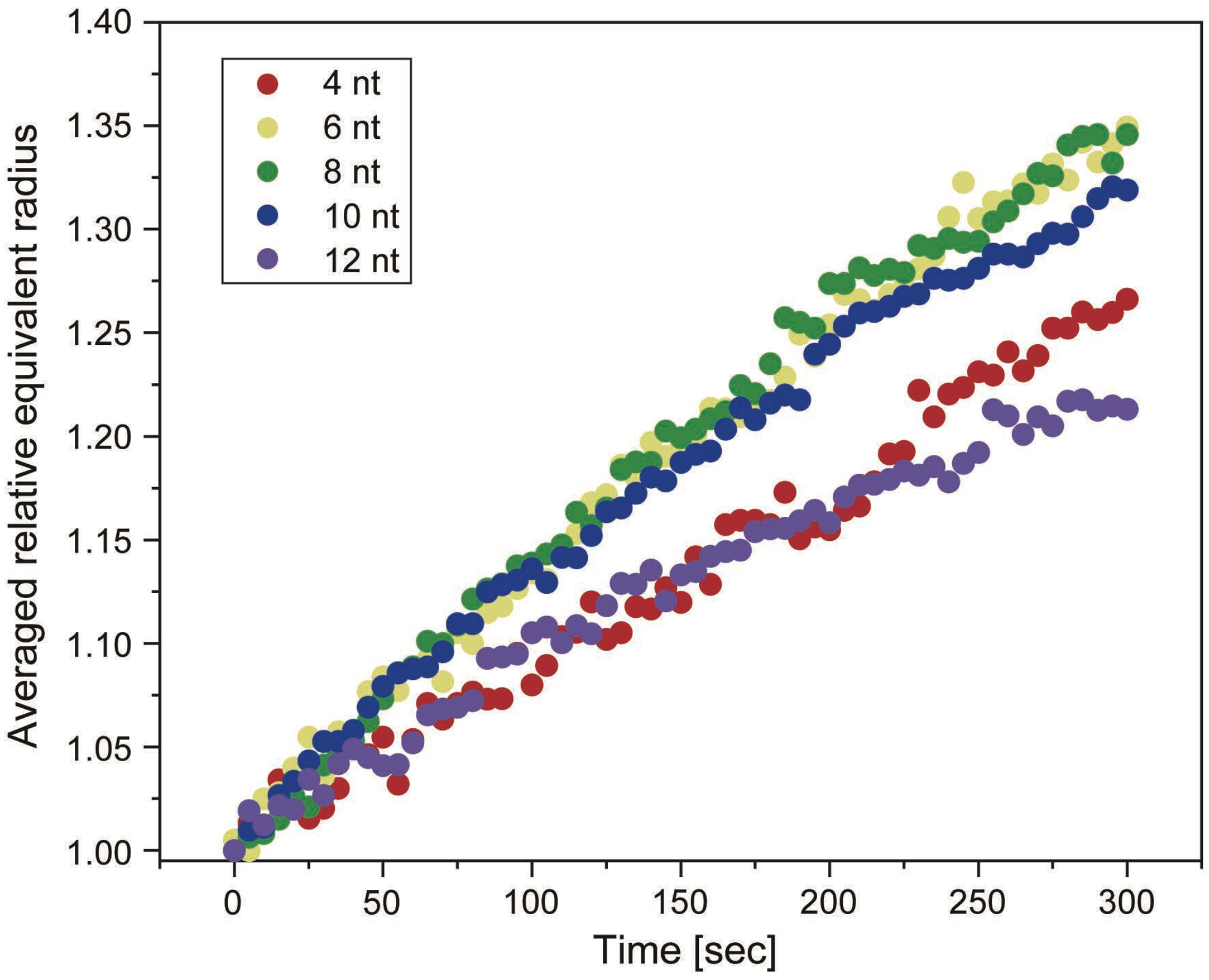

**Supplementary Fig. 2**. Time course of the size of DNA droplets composed of Y-motifs at around $T_d$ (the phase-change temperature between dispersed and droplet phases) of each sticky end (SE). To measure the averaged relative equivalent radius of the DNA droplets with different nucleotide lengths of SEs, averaged equivalent radius (AER) in each SE was calculated based on the area of the droplets visualized by a confocal microscope. At each time point, the number of droplets counted for the calculation was over 900. The AERs were then divided by the minimum value to obtain the averaged relative equivalent radius.

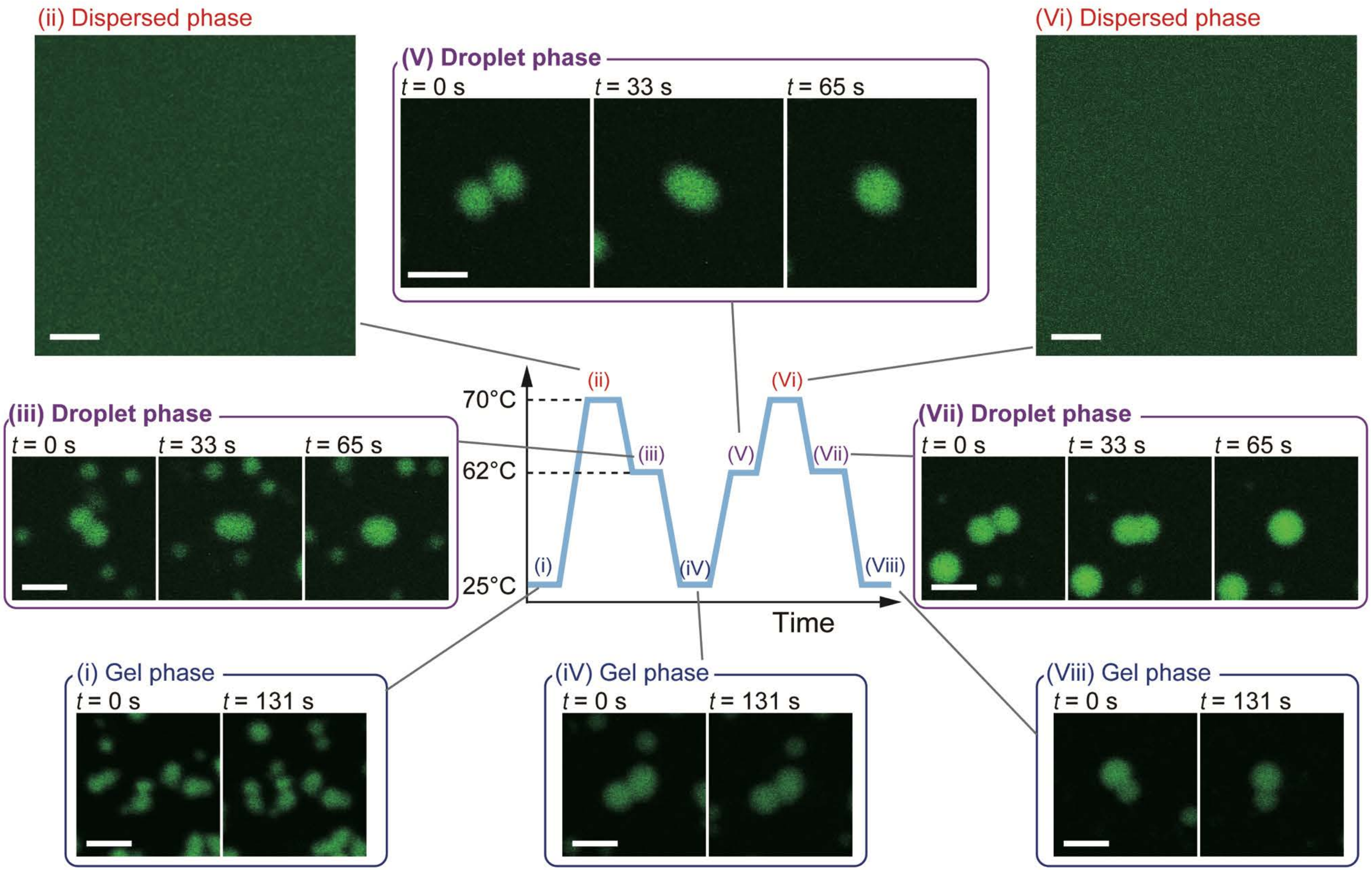

**Supplementary Fig. 3**. Repetitive phase changes of Y-motifs with 8 nucleotide long sticky ends in response to temperature changes. The temperature was changed at a rate of ± 20 °C/min. Scale bars: (ii) and (Vi), 30 μm; all others, 10 μm.

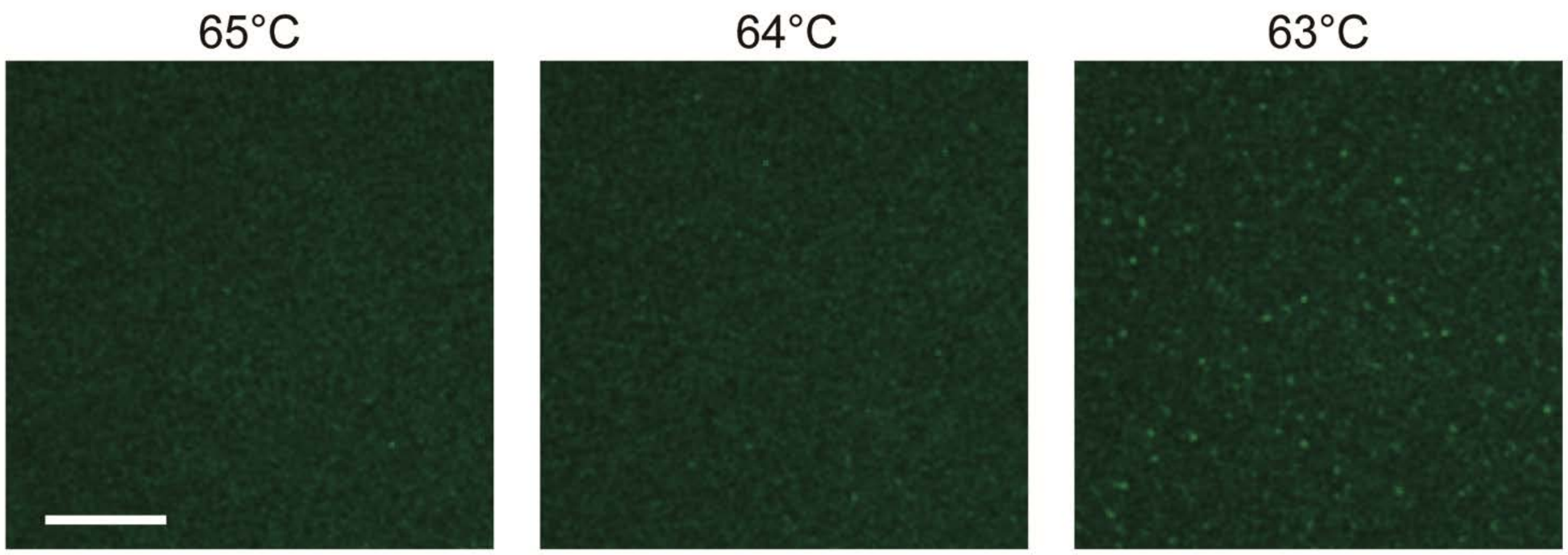

**Supplementary Fig. 4**. Microscopy images representing the moment when DNA droplets composed of Y-motifs with 8 nucleotide long sticky ends are formed by phase separation in response to decreases in temperature. Scale bar: 50 µm.

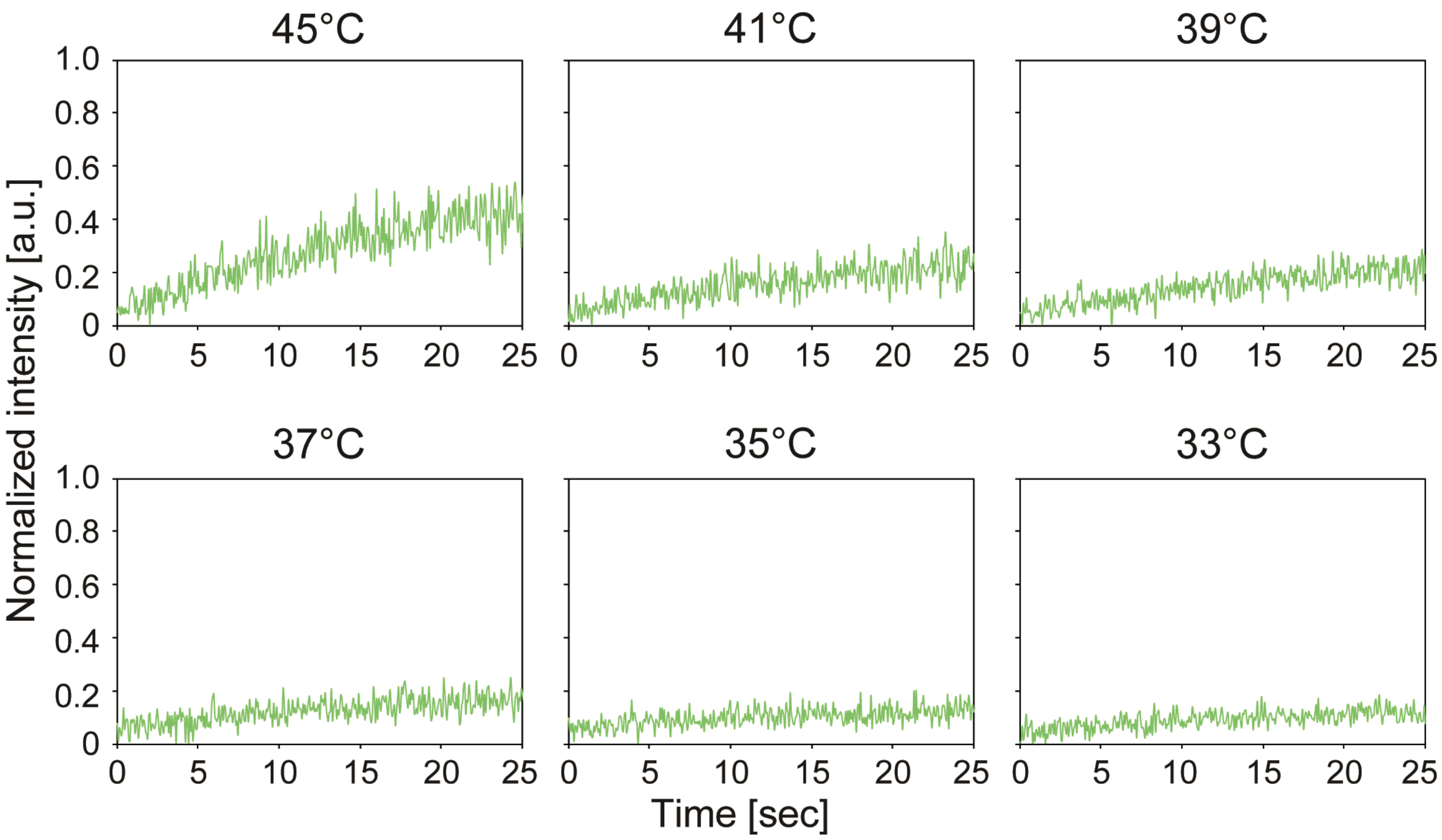

**Supplementary Fig. 5**. Representative results for the FRAP experiments with Y-motifs with 8 nucleotide long sticky ends. The half region of droplets was photo-bleached, and the changes in fluorescence intensity were measured in the bleached regions at each temperature.

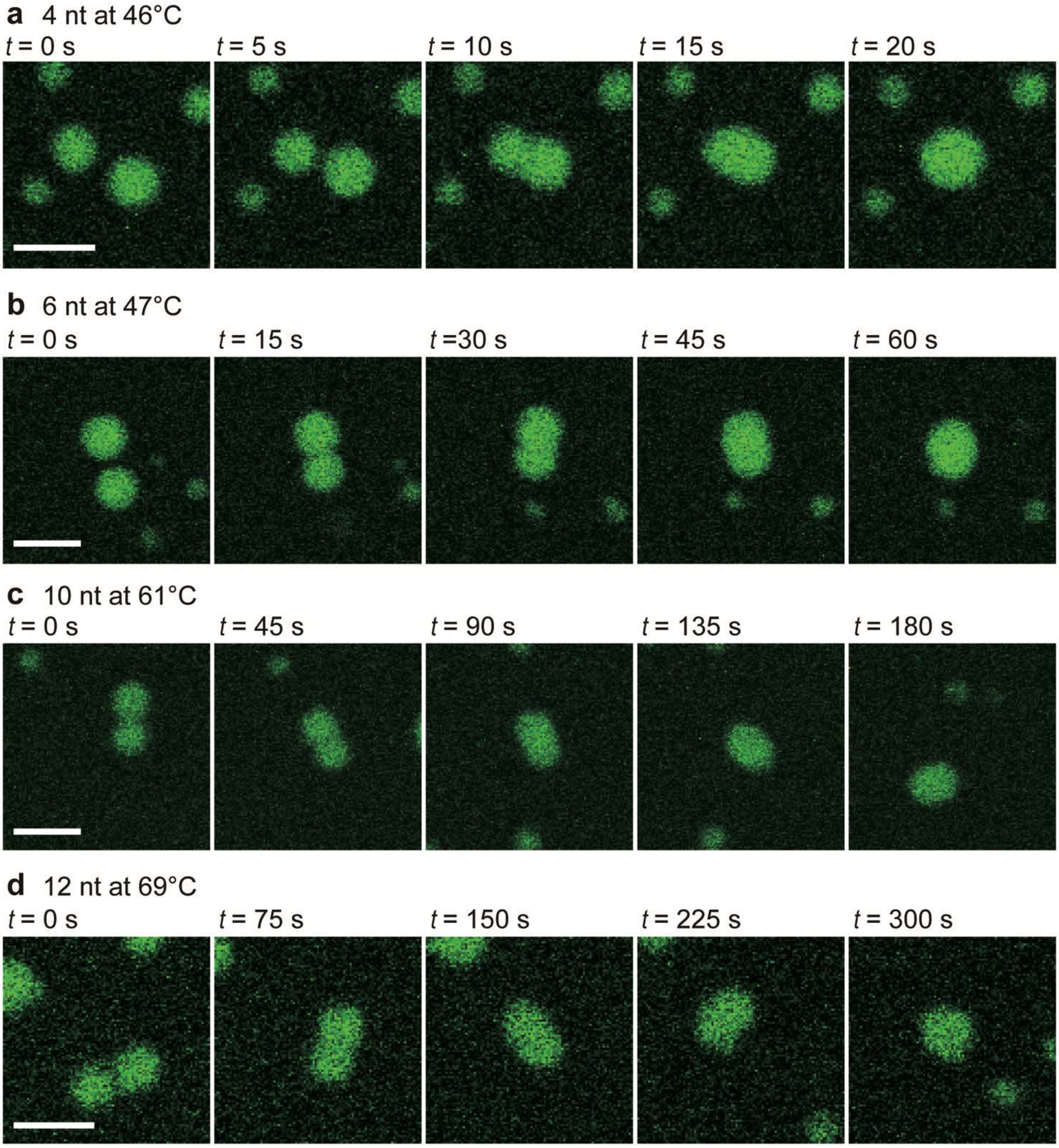

**Supplementary Fig. 6**. Fusion events of DNA droplets with Y-motifs that have 4 (**a**), 6 (**b**), 10 (**c**), and 12 nucleotide (**d**) long sticky ends. Scale bars: 10 μm.

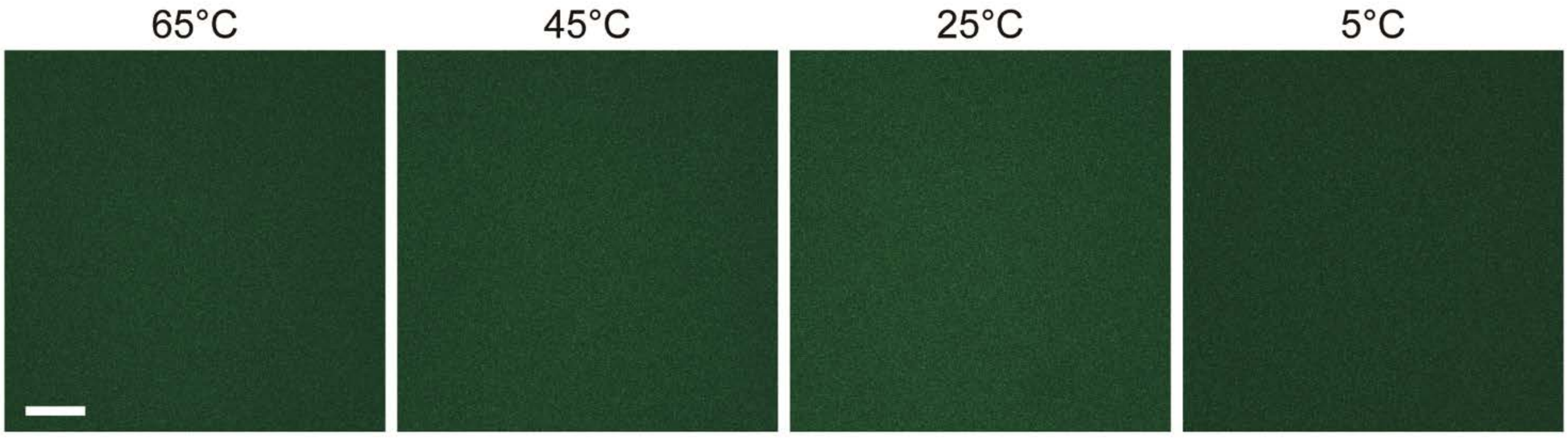

**Supplementary Fig. 7**. Microscopy images for Y-motifs with 2 nucleotide long sticky ends visualized at the temperatures shown above the images. The sample solution in the observation chamber was incubated at 85 °C for 3 min; then the temperature was decreased to 5 °C at a rate of –1 °C/min. During the annealing process, no micrometre-sized structures were observed. Scale bar: 100 μm.

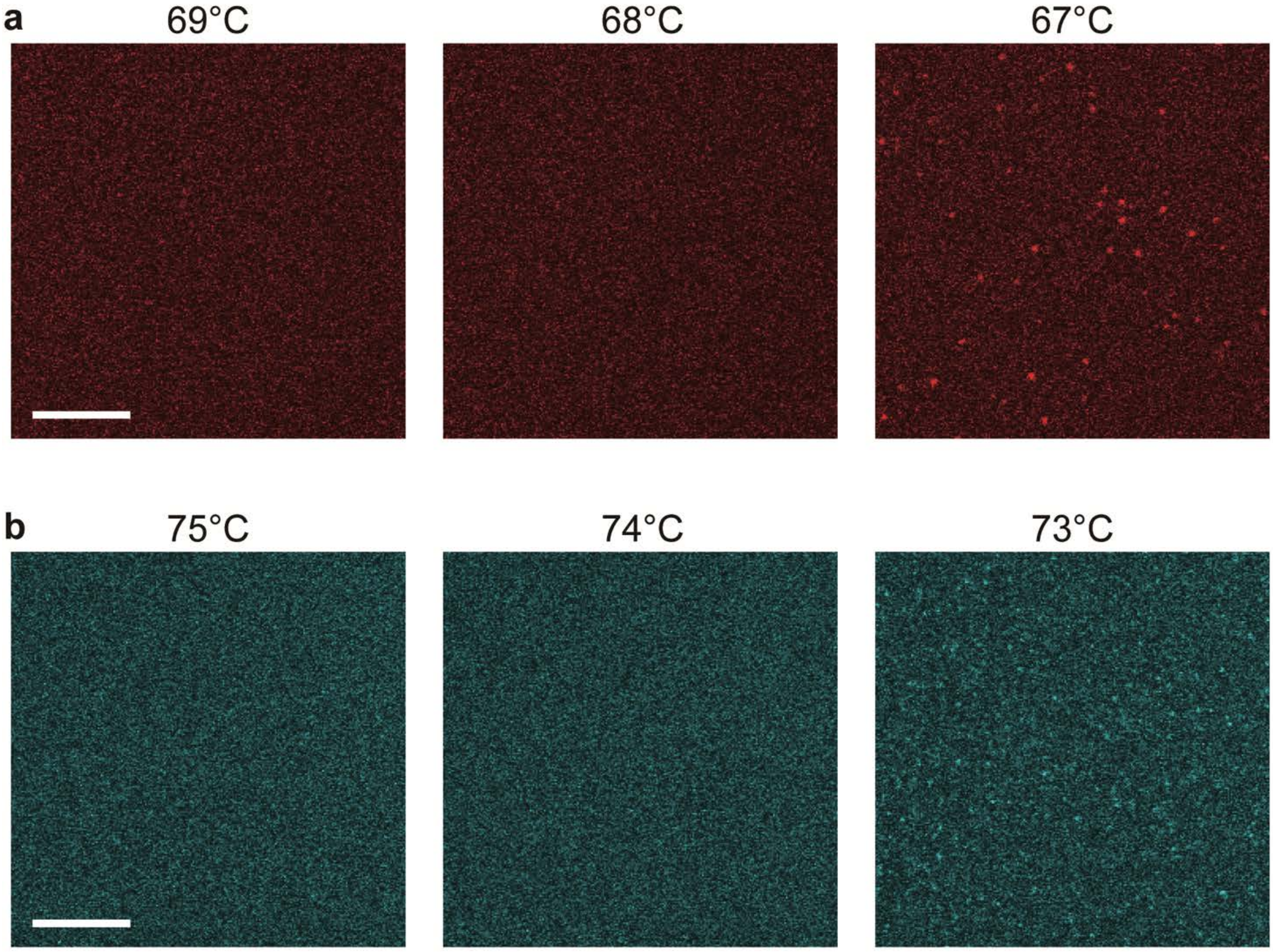

**Supplementary Fig. 8**. Microscopy images representing the moment when DNA droplets composed of motifs having four (**a**) and six branches with 8 nucleotide long sticky ends (**b**) are formed by phase separation in response to temperature decreases. Scale bars: 50 μm.

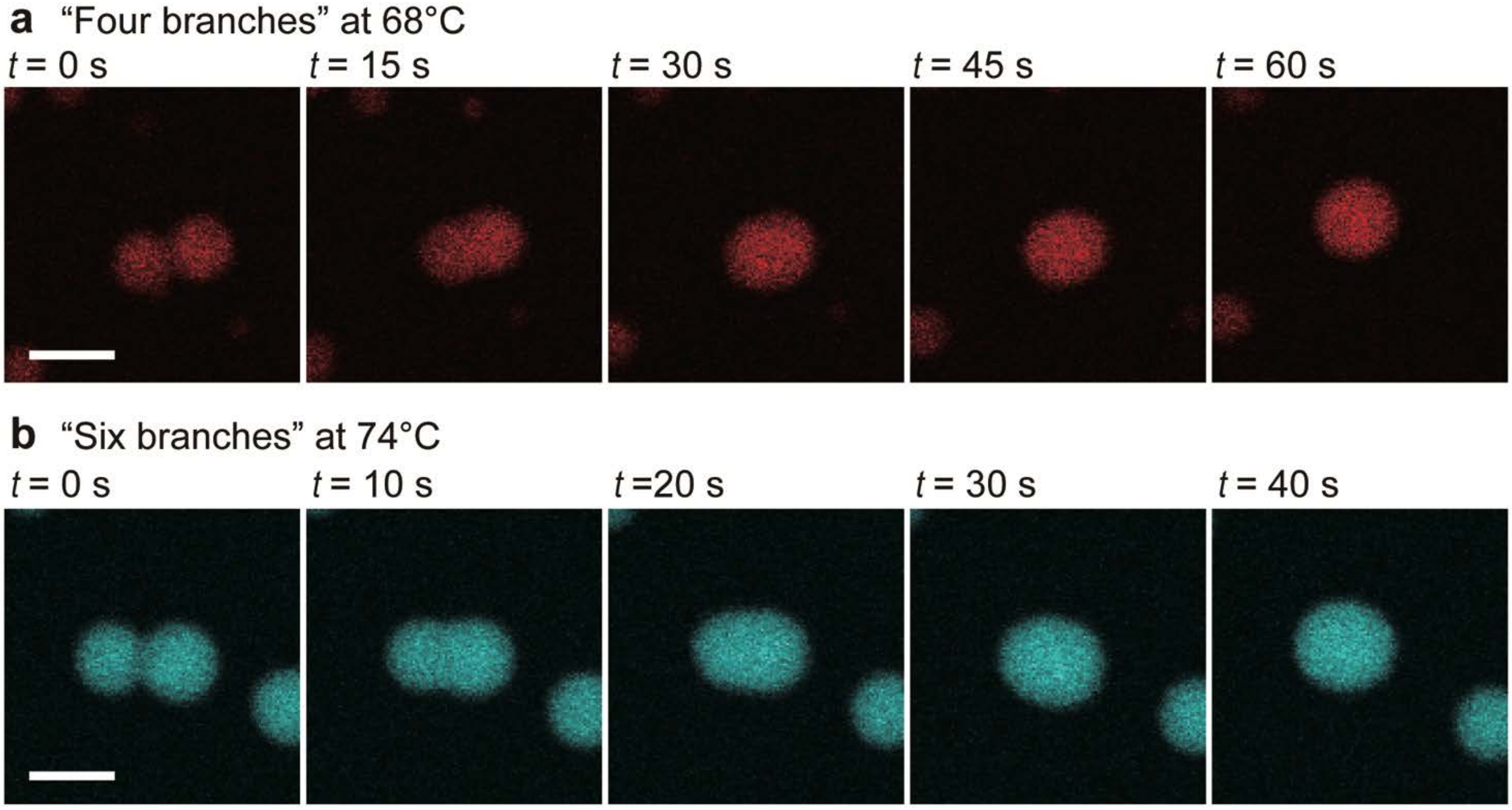

**Supplementary Fig. 9**. Fusion behaviour of DNA droplets composed of motifs having four (**a**) and six (**b**) branches with 8 nucleotide long sticky ends. Scale bars: 10 μm.

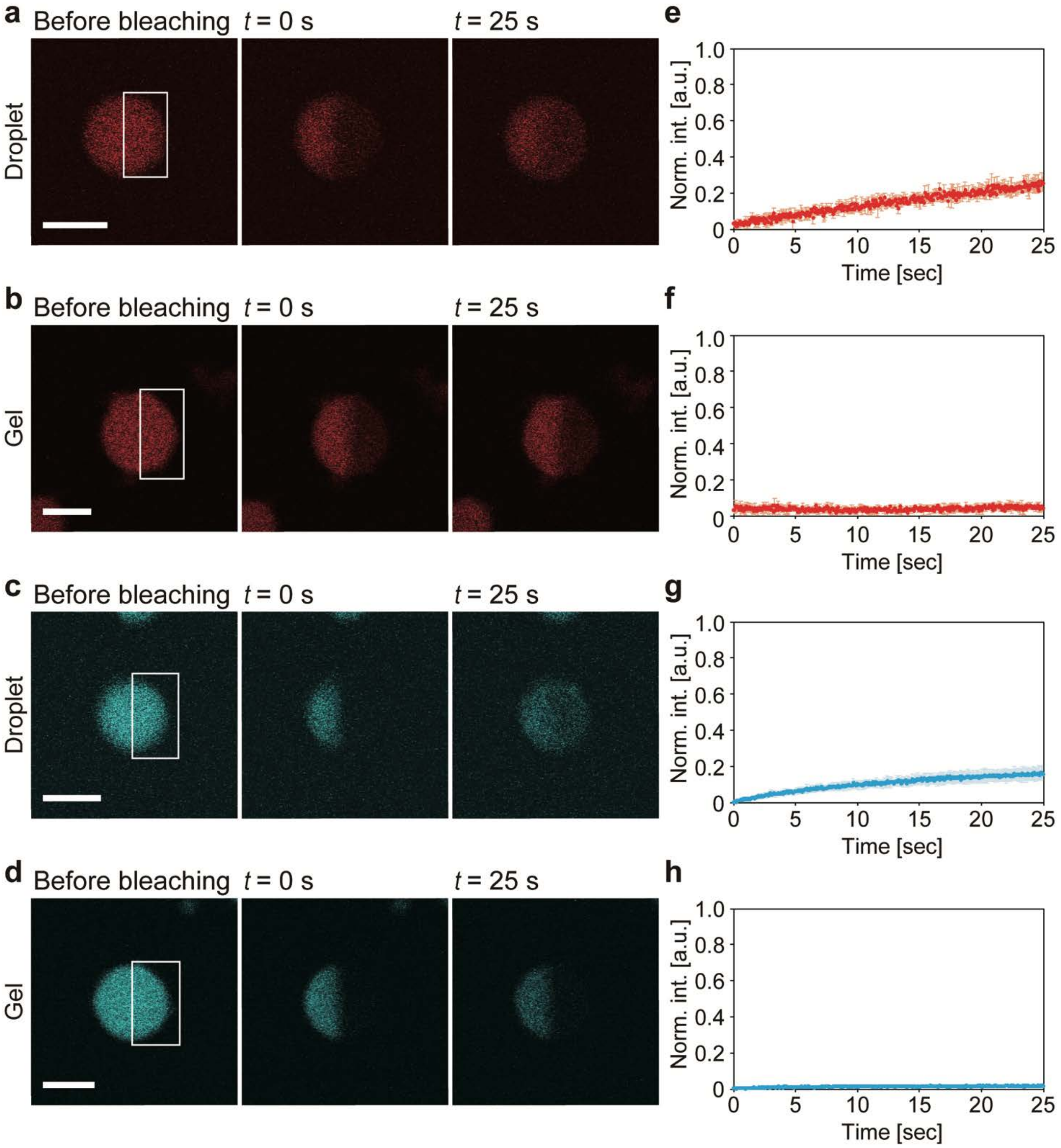

**Supplementary Fig. 10**. FRAP experiments for DNA droplets and hydrogels composed of motifs with four or six branches with 8 nucleotide long sticky ends. **a** and **b**, Microscopy images of FRAP experiments for DNA droplets (**a**) and hydrogels (**b**) composed of motifs with four branches with 8 nucleotides long sticky ends visualized at 68 and 52 °C, respectively. **c** and **d**, Microscopy images in FRAP experiments for DNA droplets (**c**) and hydrogels (**d**) composed of motifs with six branches with 8 nucleotide sticky ends, visualized at 74 and 60 °C, respectively. The bleached region is indicated by a white box in each image. Scale bars: 10 μm. **e, f, g,** and **h** are time series of fluorescence intensity corresponding to the images shown to the left of each graph. Error bars means the mean ± standard deviation (S.D.). The number of measurements ($n$) = 3 .

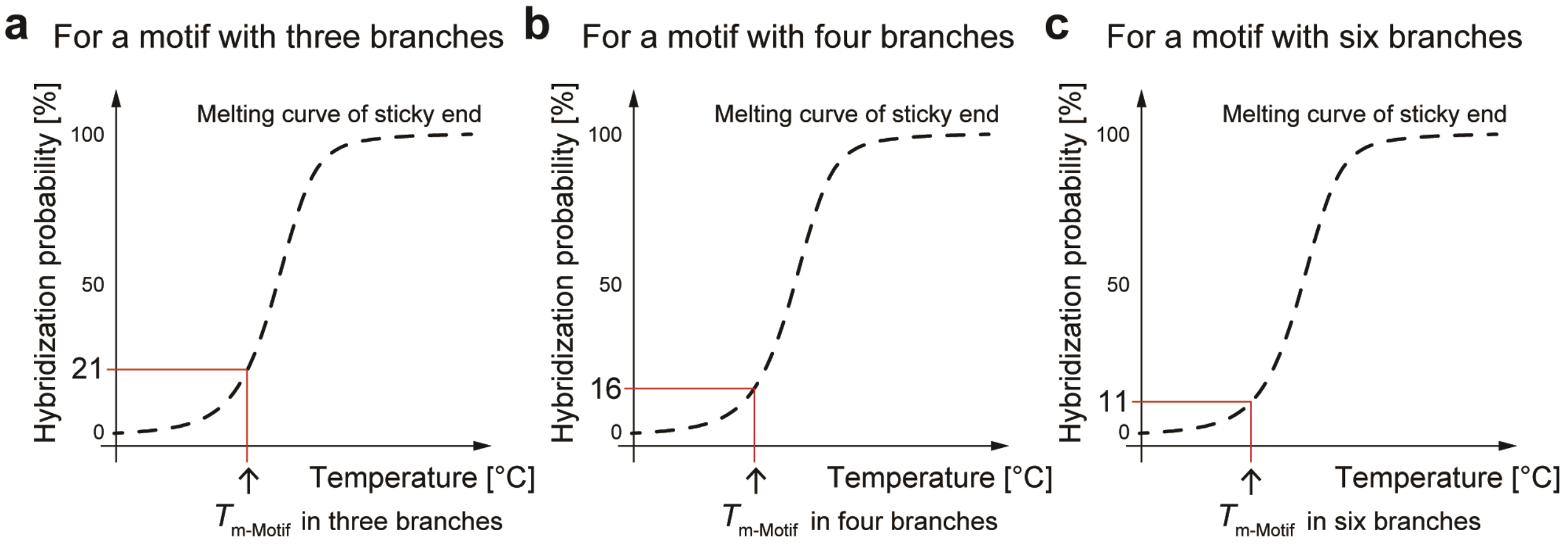

**Supplementary Fig. 11**. Definition of $T_{\text{m-Motif}}$ that represents the temperature at which at least one of the multiple sticky ends (SEs) can hybridize at a 50 % ratio. To satisfy this, the hybridization probability of SEs in a motif with three, four, and six branches were 21, 16, 11 %, respectively.

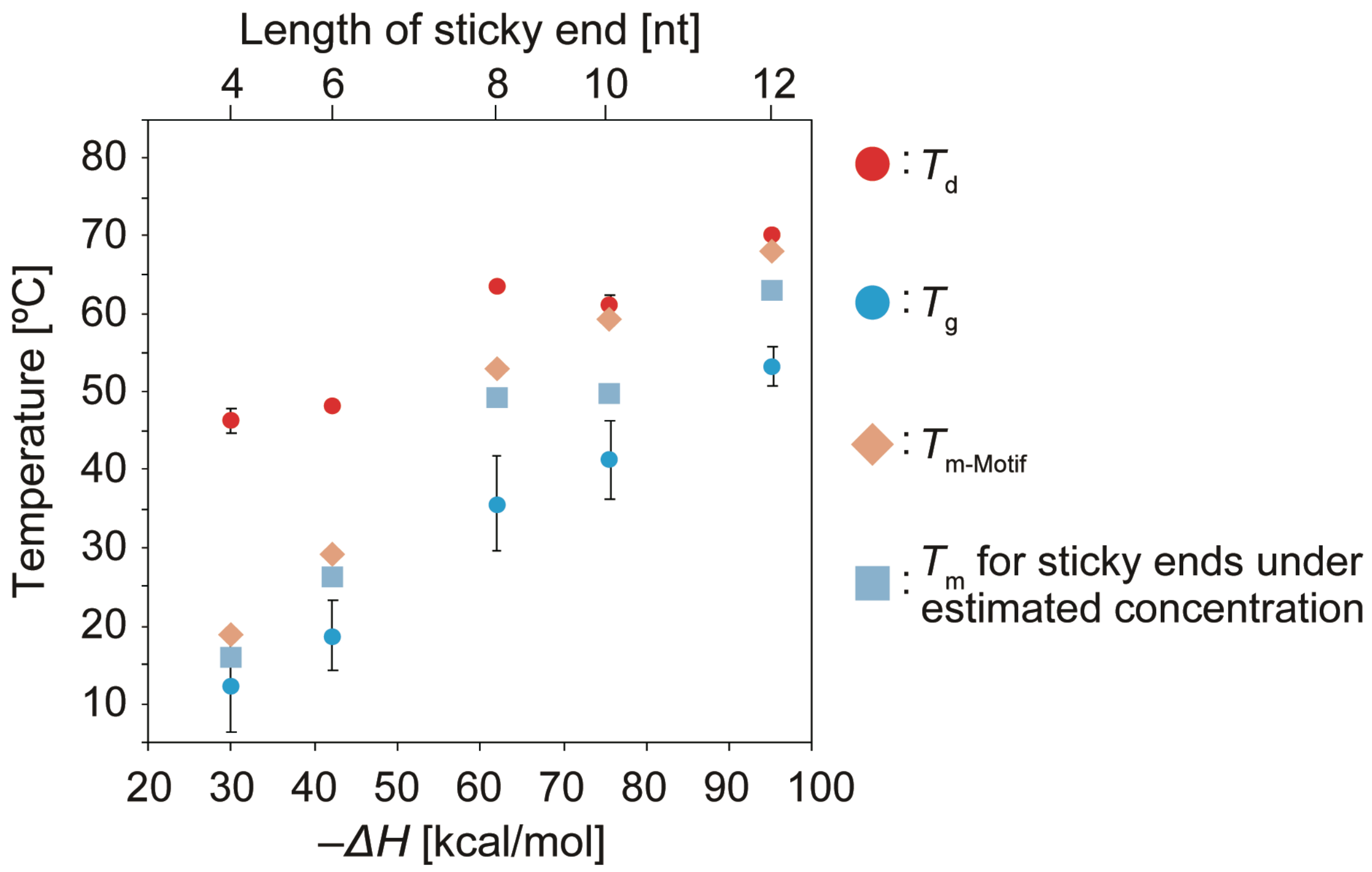

**Supplementary Fig. 12**. Comparison of $T_d$ (the phase-change temperature between dispersed and droplet phases), $T_g$ (the phase-change temperature between the droplet and gel phases), $T_{m\text{-Motif}}$ (the $T_m$ for the motifs at which half of the motifs in a solution are connected), and $T_m$ for the sticky ends with the experimentally estimated concentrations in the Y-motif with each length of sticky end. Error bars indicate standard deviation (mean ± S.D., $n = 3$).

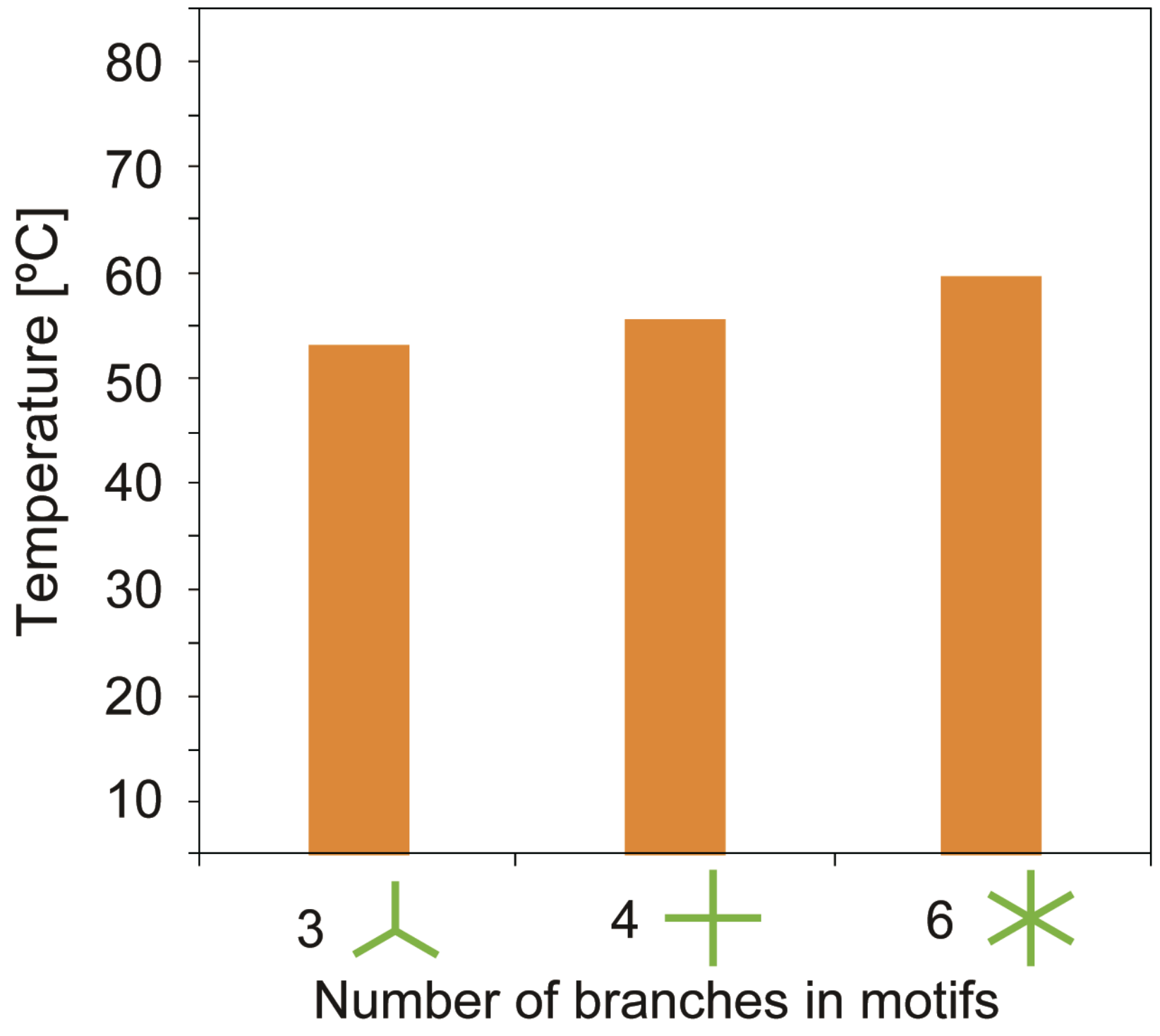

**Supplementary Fig. 13**. $T_{\text{m-Motif}}$ (the $T_m$ for the motifs at which half of the motifs in a solution are connected) for DNA droplets with three, four, and six, 8 nucleotide long sticky ends.

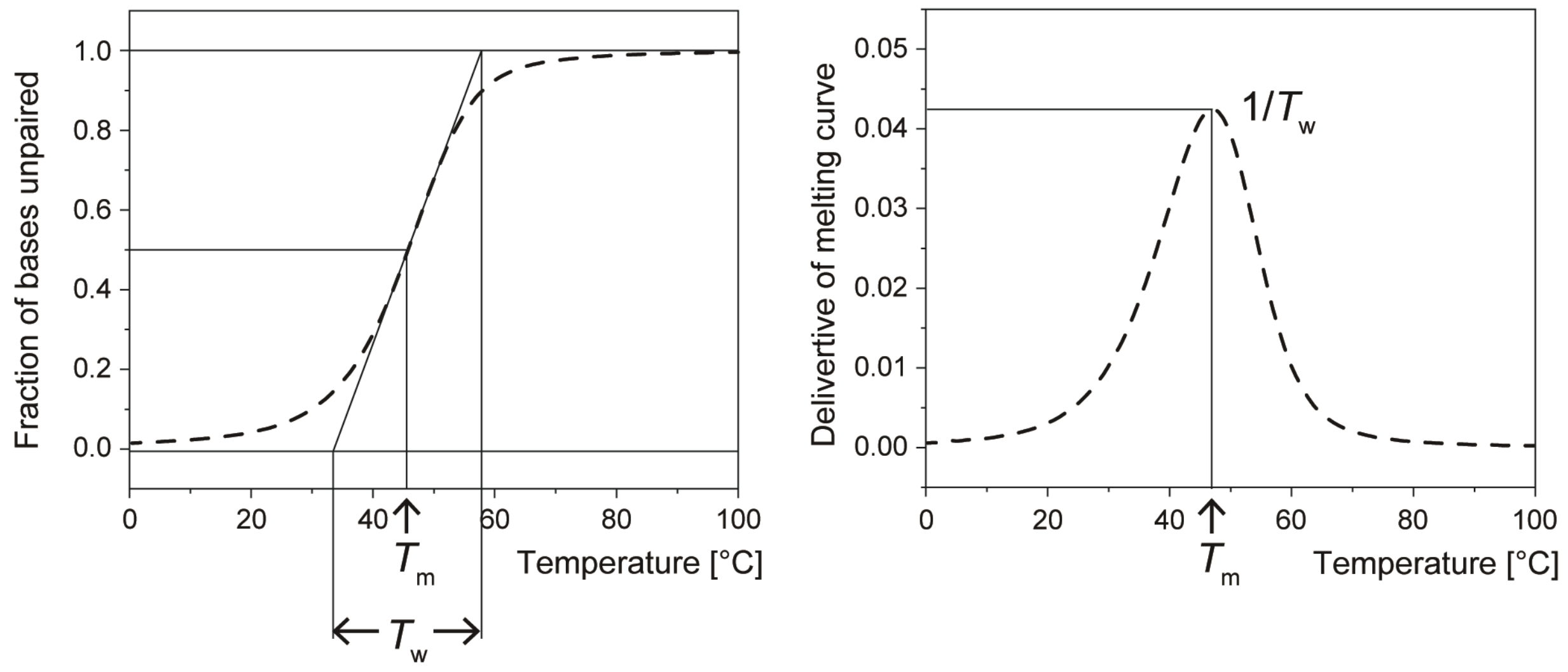

**Supplementary Fig. 14**. Definition of the width of the melting range ($T_w$). $T_w$ was determined as the distance between intersections with horizontal lines 'Fraction of bases unpaired' = 0 and 1 of the tangent in a melting curve at the $T_m$.

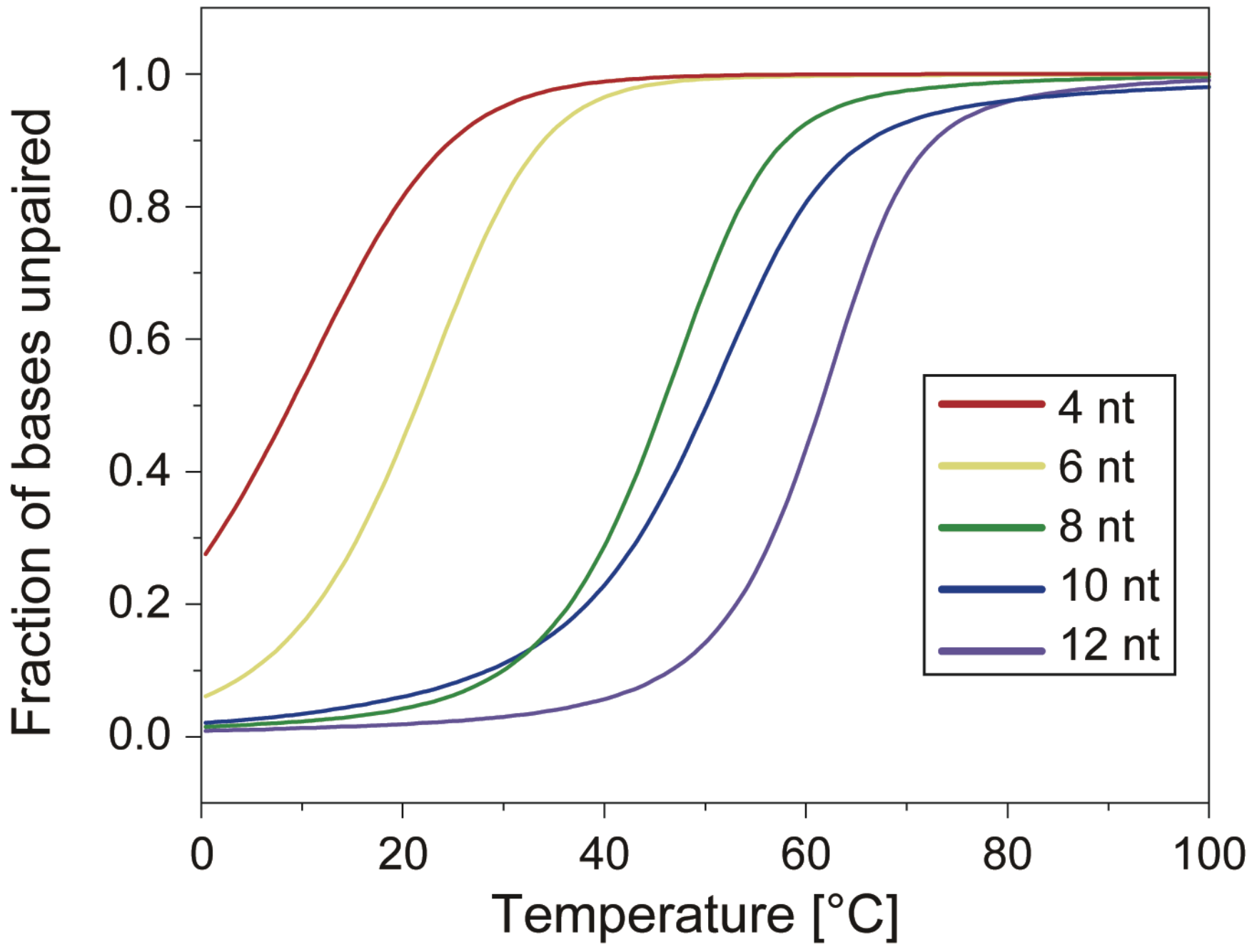

**Supplementary Fig. 15**. Melting curves for sticky ends that are 4, 6, 8, 10, and 12 nucleotides in length obtained through numerical simulation using NUPACK software[1].

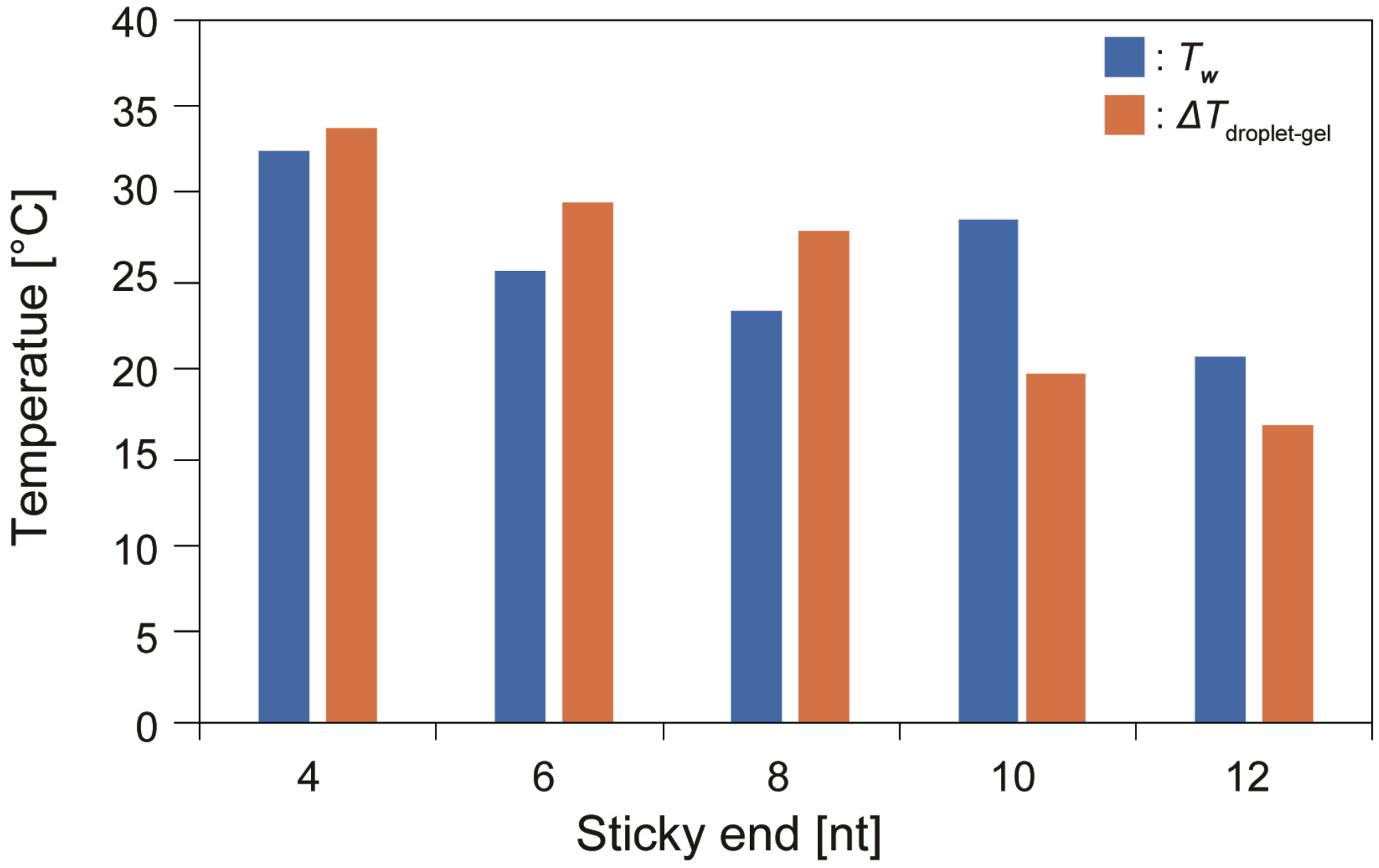

**Supplementary Fig. 16**. Temperature differences between the $T_d$ (the phase-change temperature between dispersed and droplet phases) and $T_g$ (the phase-change temperature between the droplet and gel phases) ($\Delta T_{droplet\text{-}gel}$) and $T_w$ (the width of the melting range) for the different nucleotide lengths of the sticky ends examined.

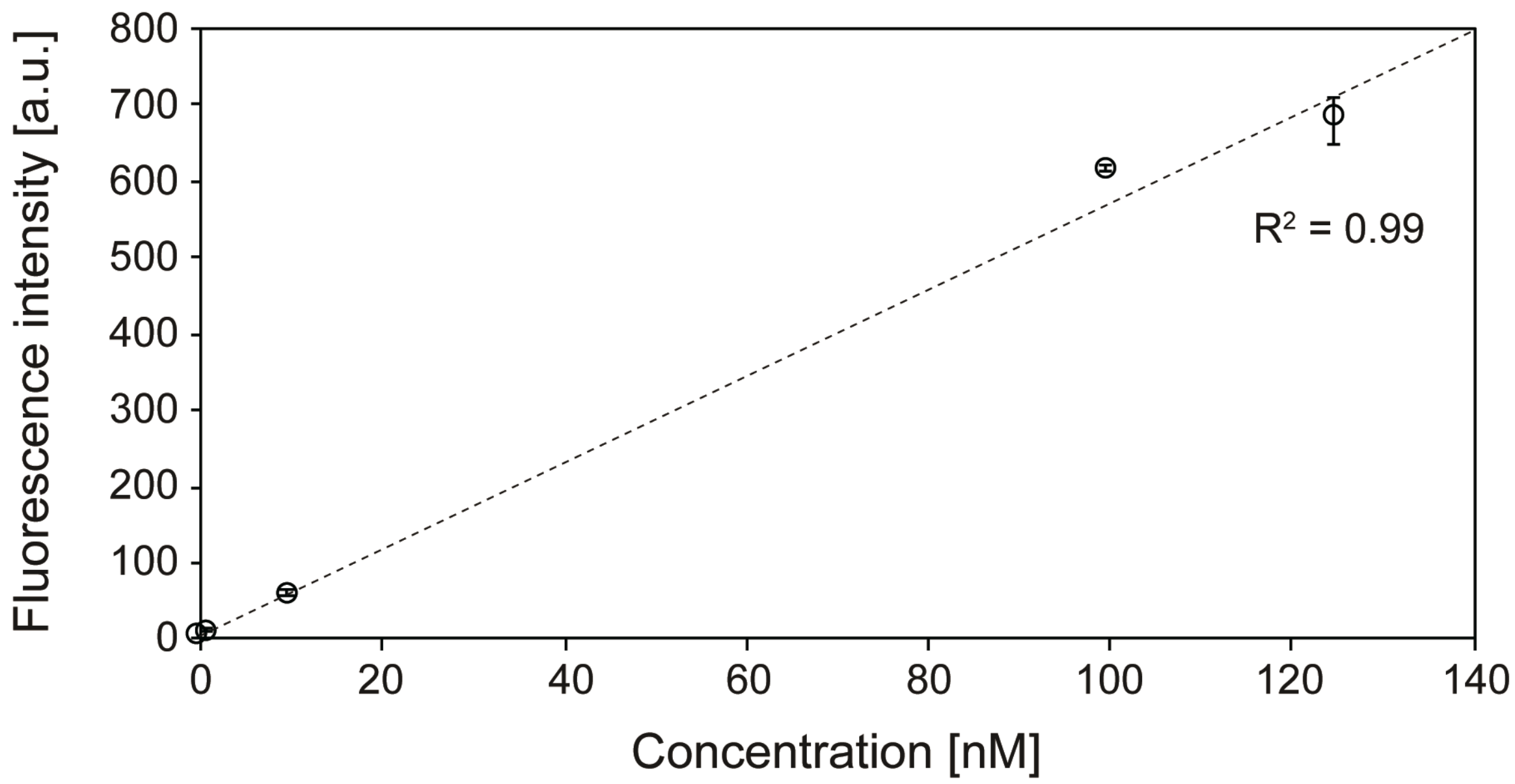

**Supplementary Fig. 17**. Calibration curve to estimate the concentration of Y-motifs in the droplet or gel phases. The dashed line through the centre of the graph was obtained through linear fitting. $R^2$ means deterministic coefficient. Error bars indicate standard deviation (mean ± S.D., $n$ = 3).

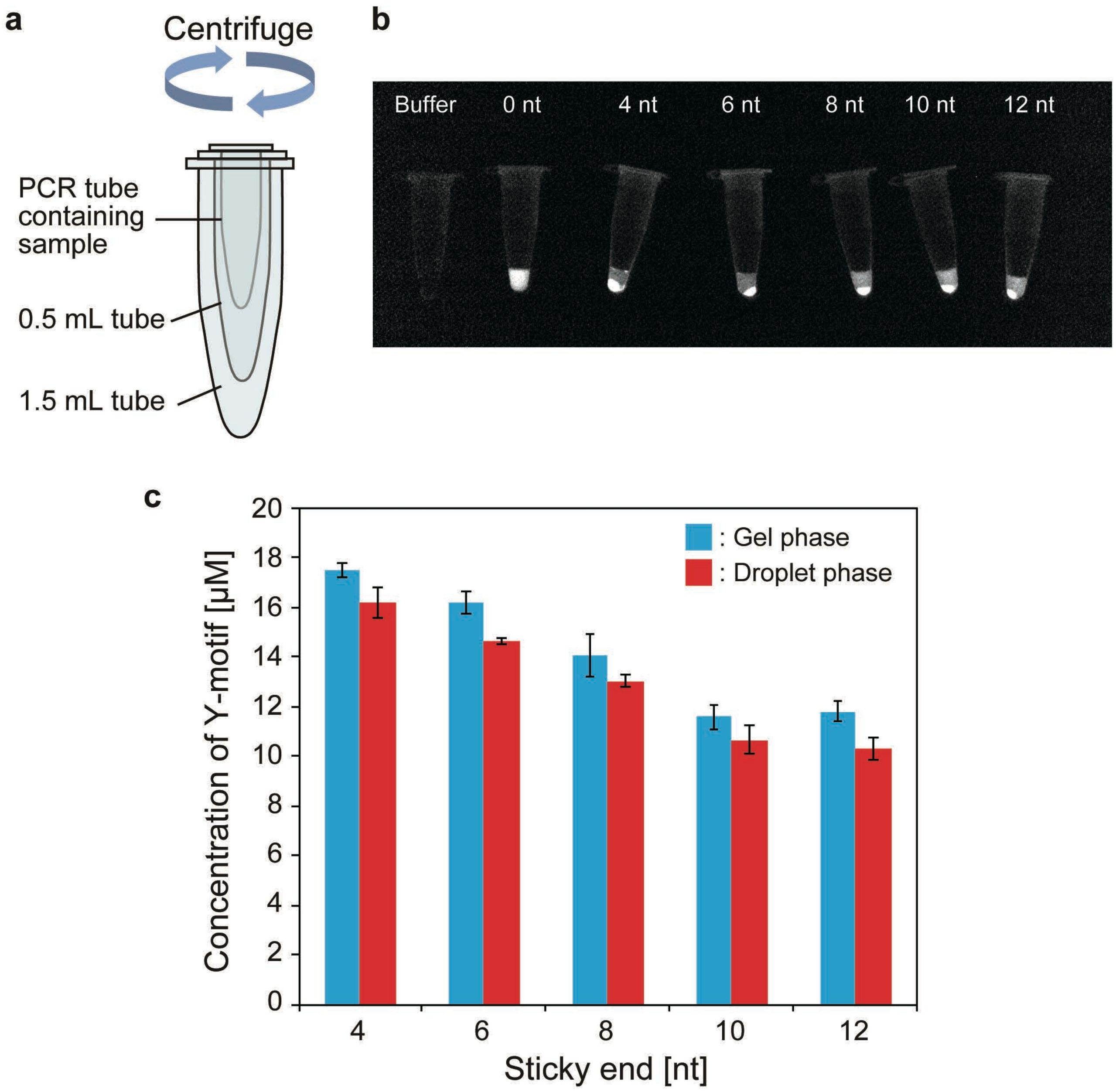

**Supplementary Fig. 18**. Estimation of Y-motif concentration for each of the different nucleotide lengths of sticky ends. **a**, Schematic illustration of the set-up. **b**, Fluorescence photo images of the PCR tubes after centrifugation. **c**, Estimates of the concentrations of Y-motifs with each length of sticky end in the droplet and gel phases. Error bars indicate standard deviation (mean ± S.D., $n$ = 3).

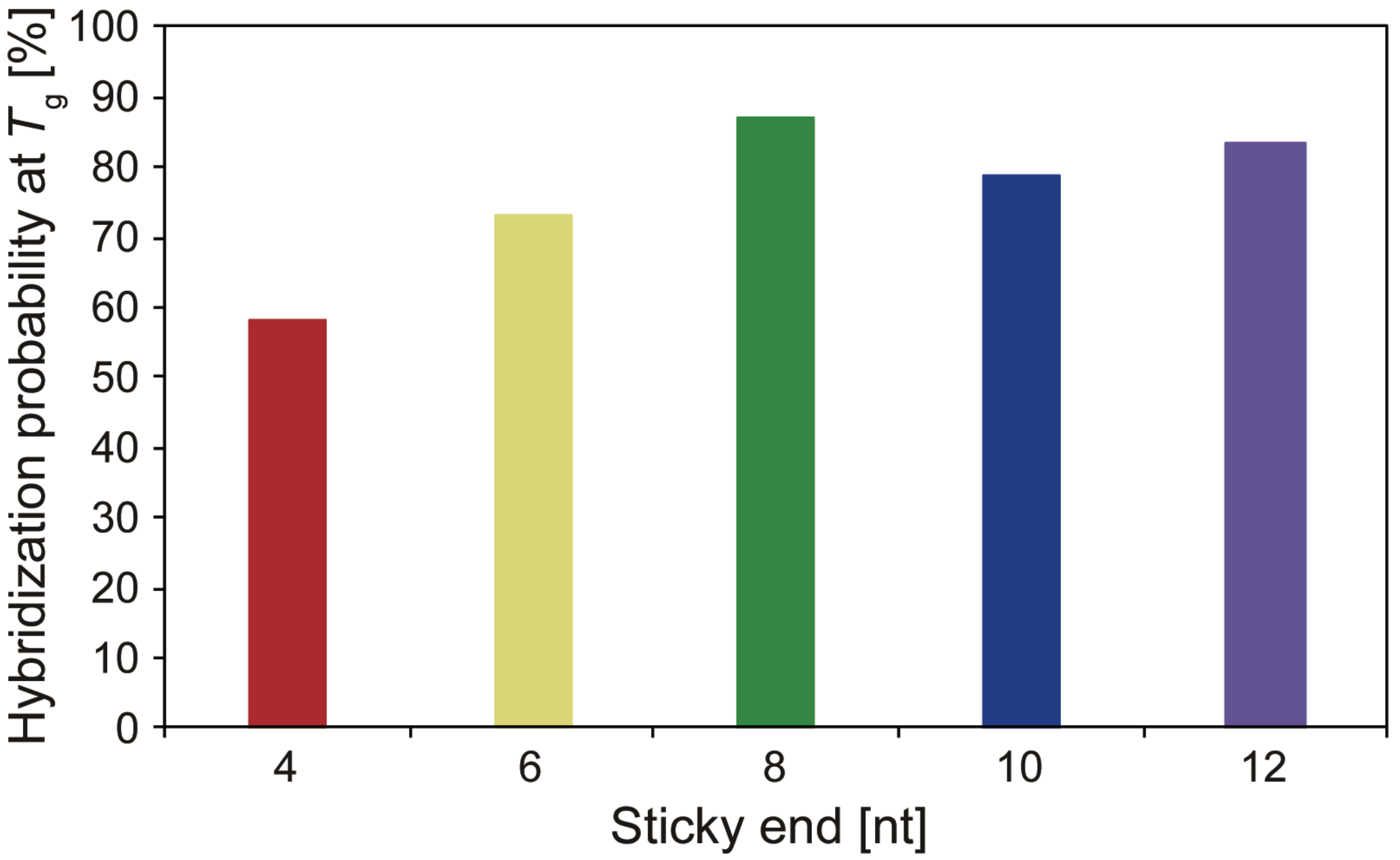

**Supplementary Fig. 19**. Hybridization probability at $T_g$ (the phase-change temperature between the droplet and gel phases) under the experimentally estimated concentrations of sticky ends. These values were obtained from numerical simulations using NUPACK software[1].

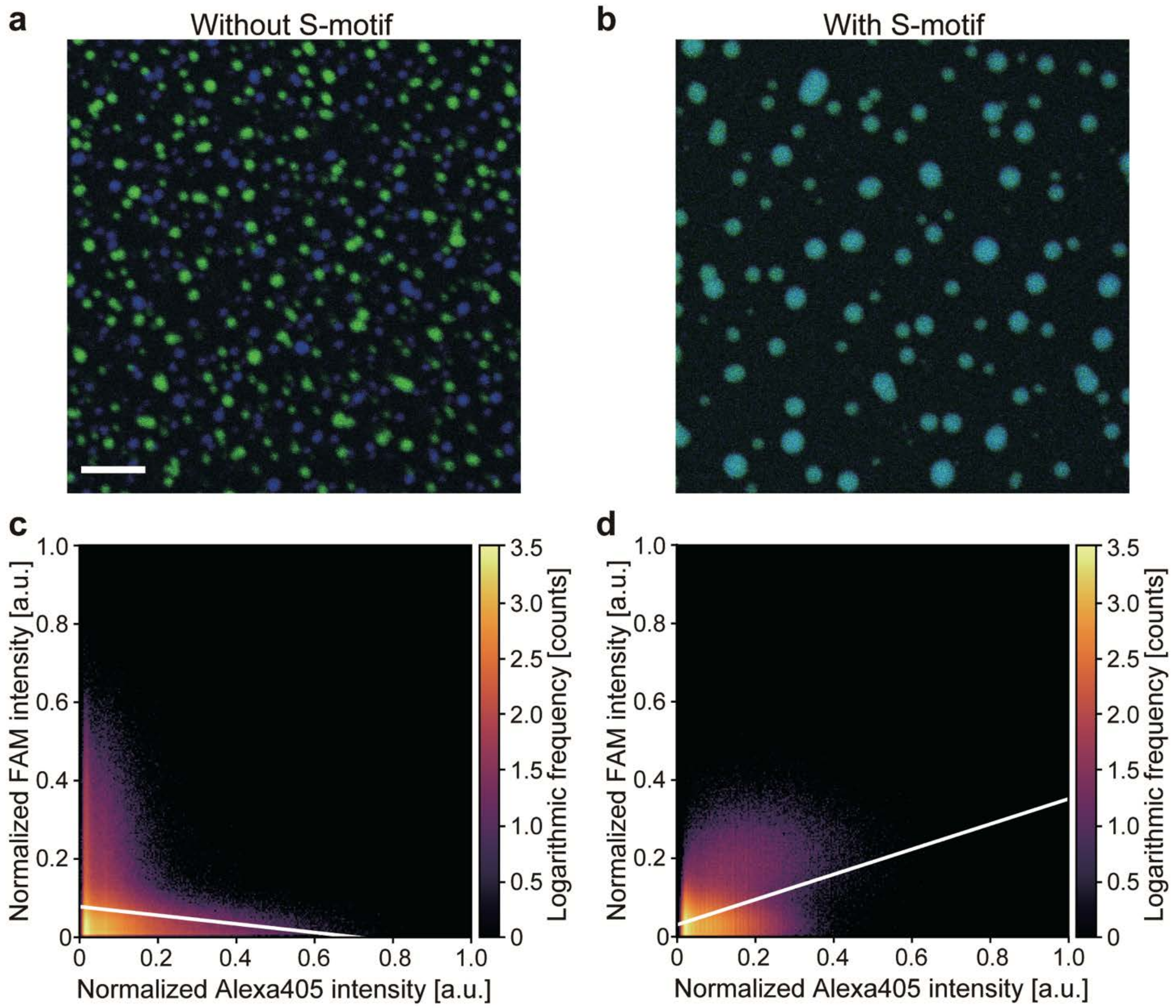

**Supplementary Fig. 20**. Colocalization analysis of FAM-labelled Y-motif and Alexa405-labelled $^{orth}$Y-motifs. **a** and **b**, Microscopy images of Y- and $^{orth}$Y-motifs in droplet phase without (**a**) and with S-motifs (**b**). Blue and green channels indicate $^{orth}$Y- (Alexa405) and Y-motifs (FAM), respectively. Scale bar: 30 μm. **c** and **d**, Two-dimensional histograms of the normalized intensity of FAM and Alexa405 without (**c**) and with S-motifs (**d**). White lines represent the linear regression line. In the graph without S-motif, the slope of the line was negative, indicating negative correlation between Y- and $^{orth}$Y-motifs (**c**). In the graph with S-motif, the slope of the line was positive, indicating positive correlation (**d**).

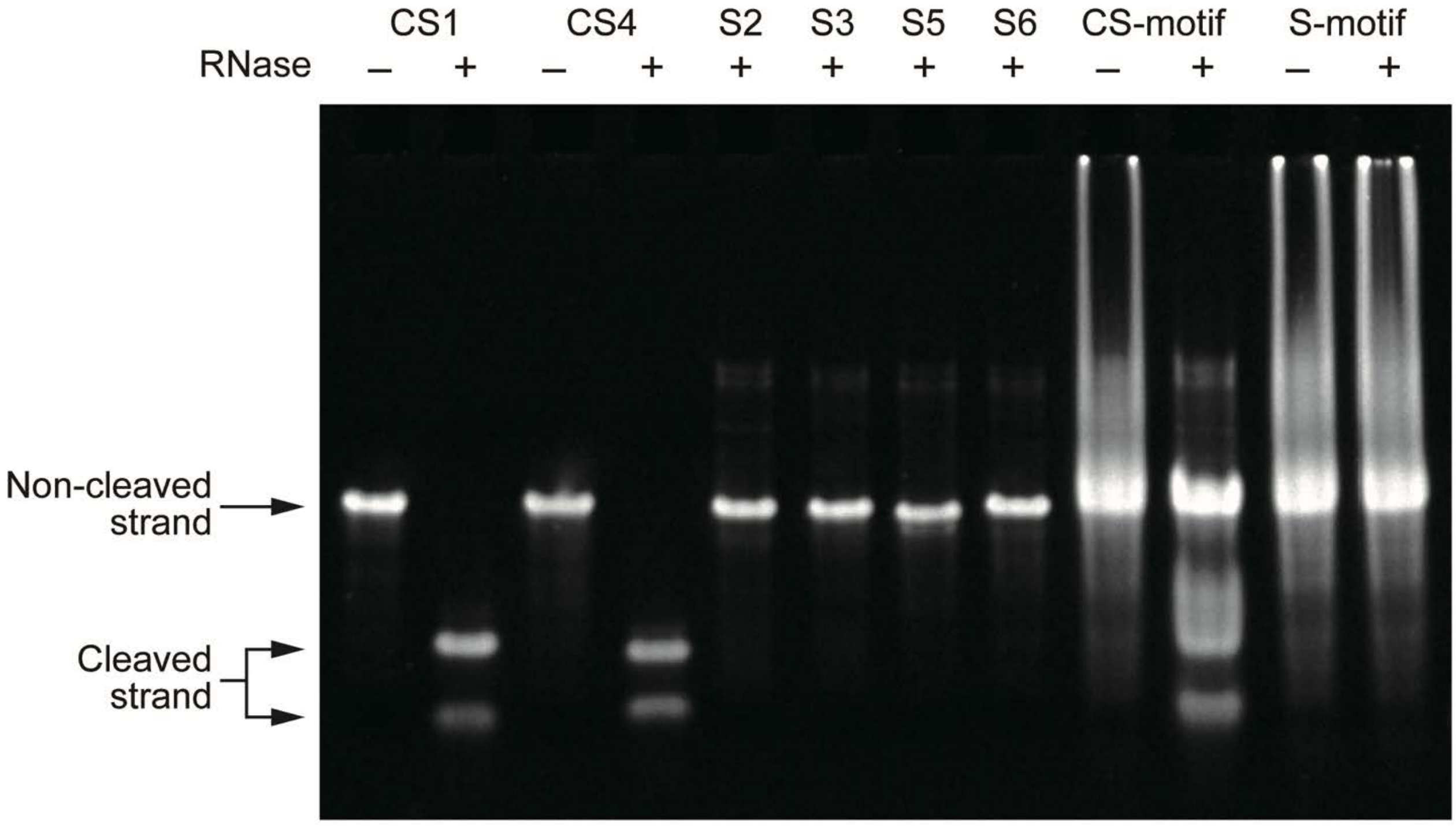

**Supplementary Fig. 21**. Denaturing gel electrophoresis of DNA-RNA chimera strands. DNA-RNA chimera strands (CS1 and CS4) were successfully cleaved with RNase A, whereas the chimera strands without RNase A and the DNA strands (S2-S5) with and without RNase were not cleaved. Similarly, the chimera strands in the CS-motifs with RNase were cleaved, whereas the cleavage was not confirmed in the CS-motif without RNase A and the S-motif with and without RNase.

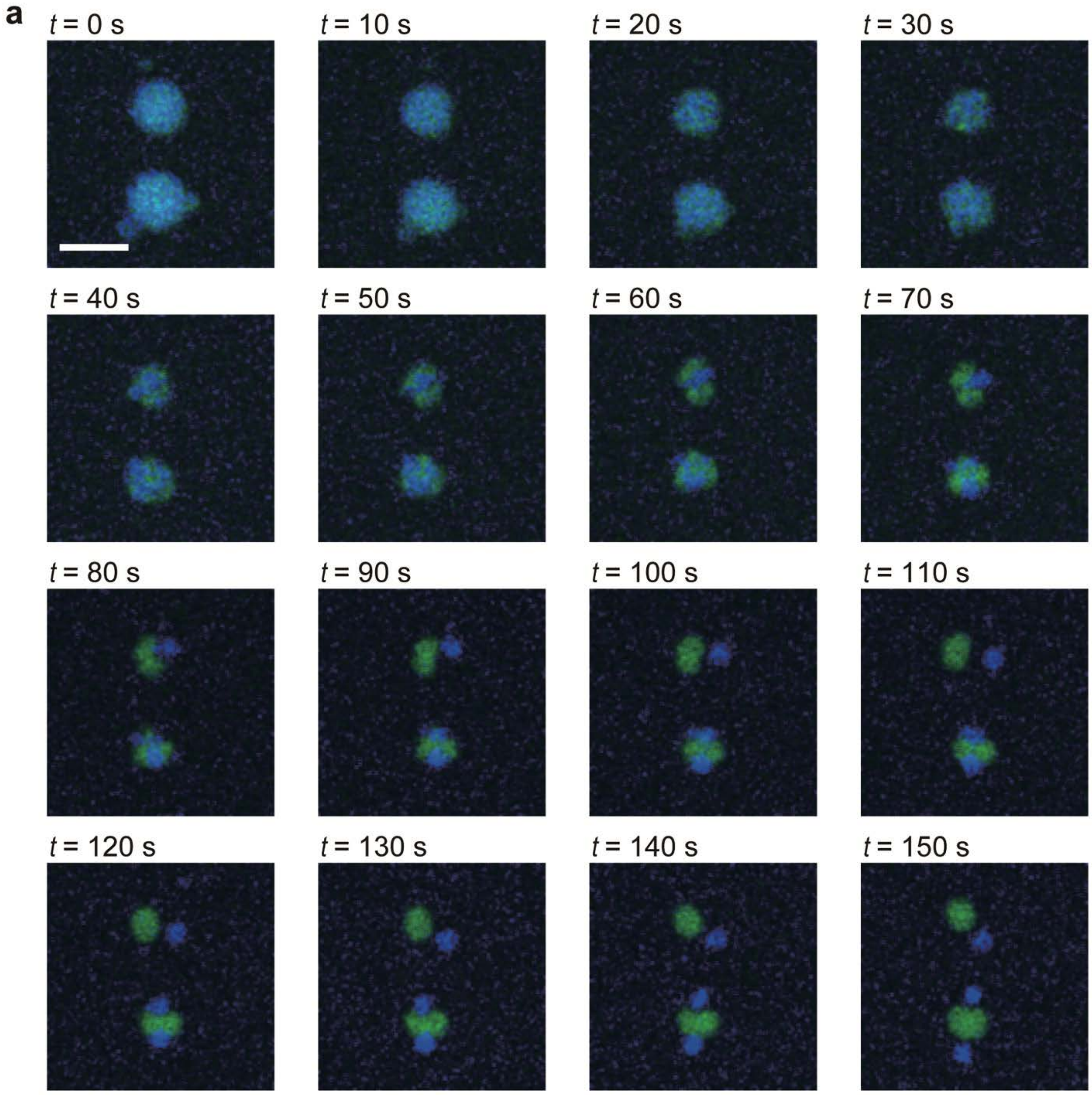

**Supplementary Fig. 22**. Detailed sequential microscopy images of the fission process of DNA droplets. Blue and green channels indicate $^{orth}$Y- and Y-motifs, respectively. Spinodal-like decomposition before the fission was observed during the fission process. We considered that the driving force for the fission is electrical repulsion between $^{orth}$Y- and Y-motifs. *t* = 0 s indicates 10 s after the addition of RNase. Scale bar: 20 μm.

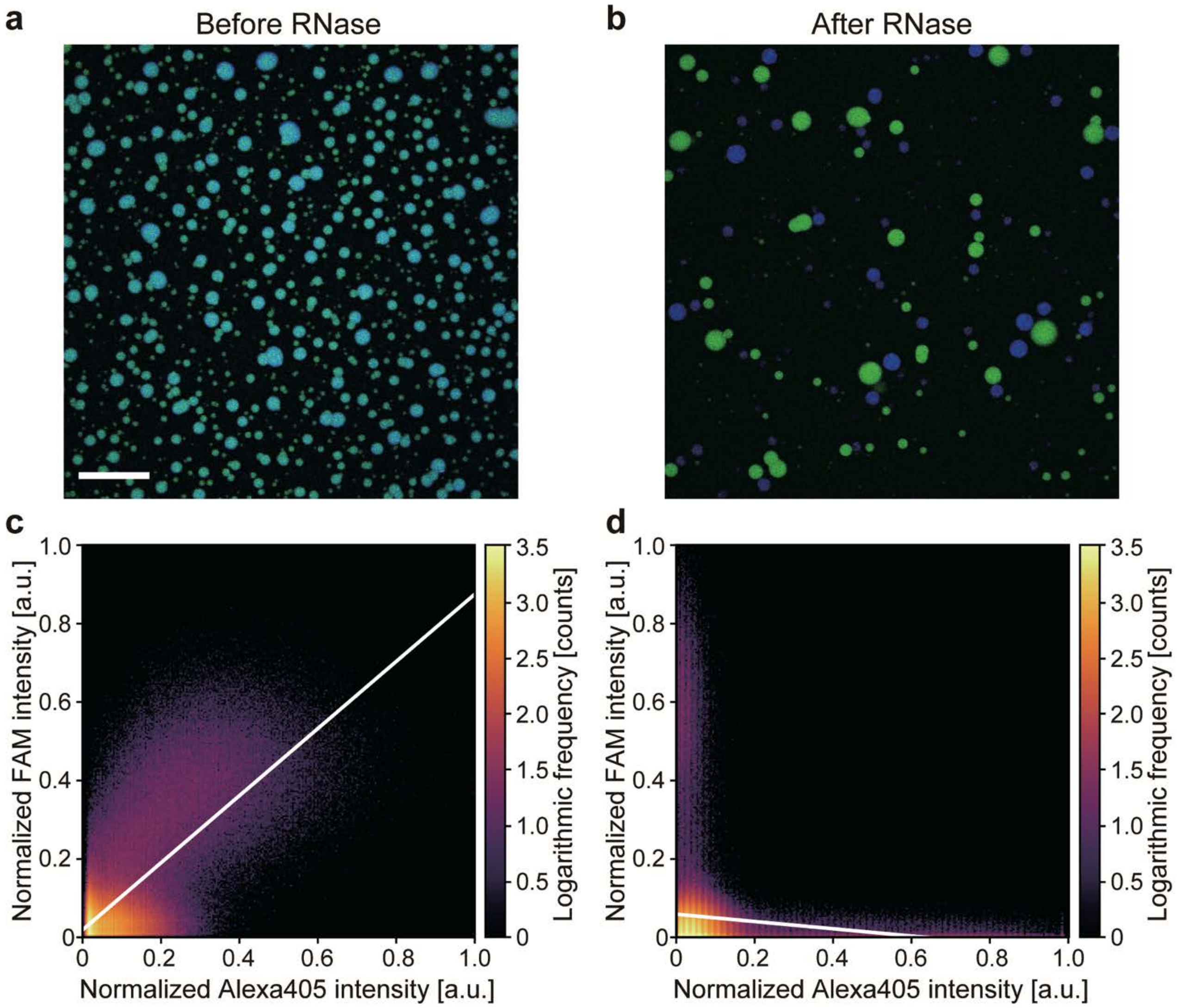

**Supplementary Fig. 23**. Colocalization analysis before and after addition of RNase A. **a** and **b**, Microscopic images of Y- and $^{orth}$Y-motifs in the droplet phase before (**a**) and after (**b**) addition of RNase A. Blue and green channels indicate $^{orth}$Y- (Alexa405) and Y-motifs (FAM), respectively. Scale bar: 100 μm. **c** and **d**, Two-dimensional histograms of the intensity of FAM and Alexa405 before (**c**) and after (**d**) addition of RNase A. White lines represent the linear regression line. Before the addition of RNase A, the slope of the line was positive, indicating positive correlation between Y- and $^{orth}$Y-motifs (**c**). After the addition, the slope of the line was negative, indicating negative correlation (**d**).

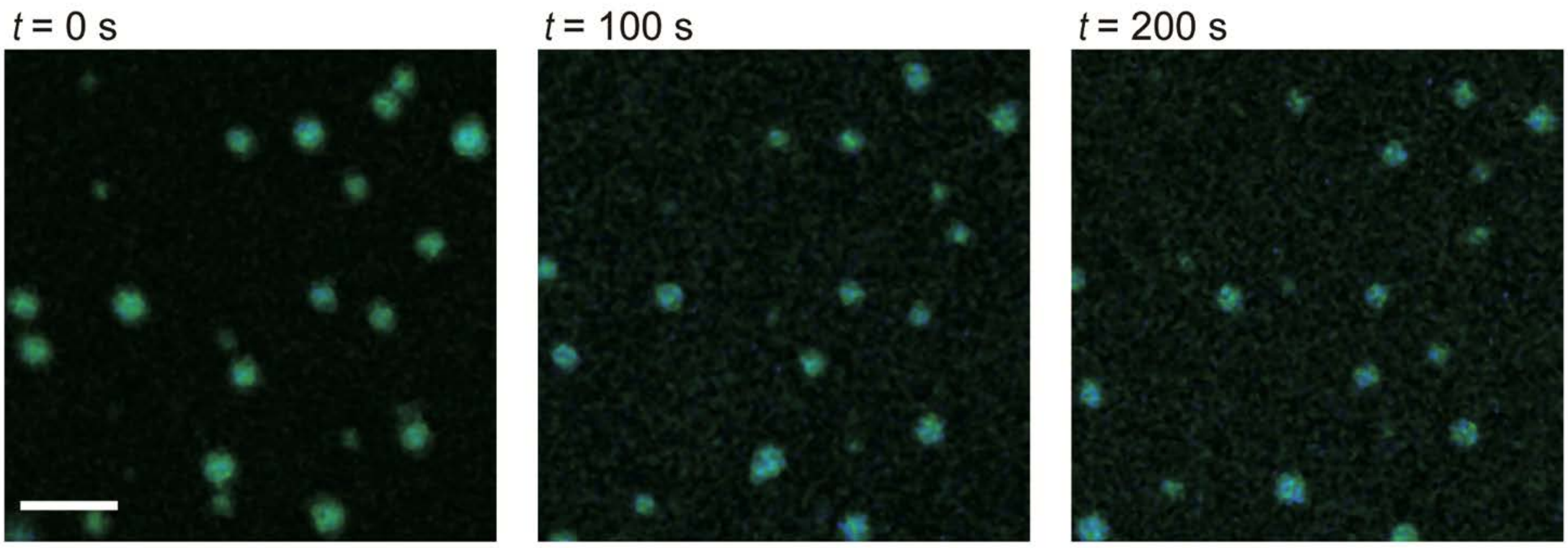

**Supplementary Fig. 24**. Droplet behaviour after addition of buffer without RNase. Blue and green channels indicate ${}^{orth}$Y- (Alexa405) and Y-motifs (FAM), respectively. $t$ = 0 s indicates 10 s after the addition of RNase. Scale bar: 20 μm.

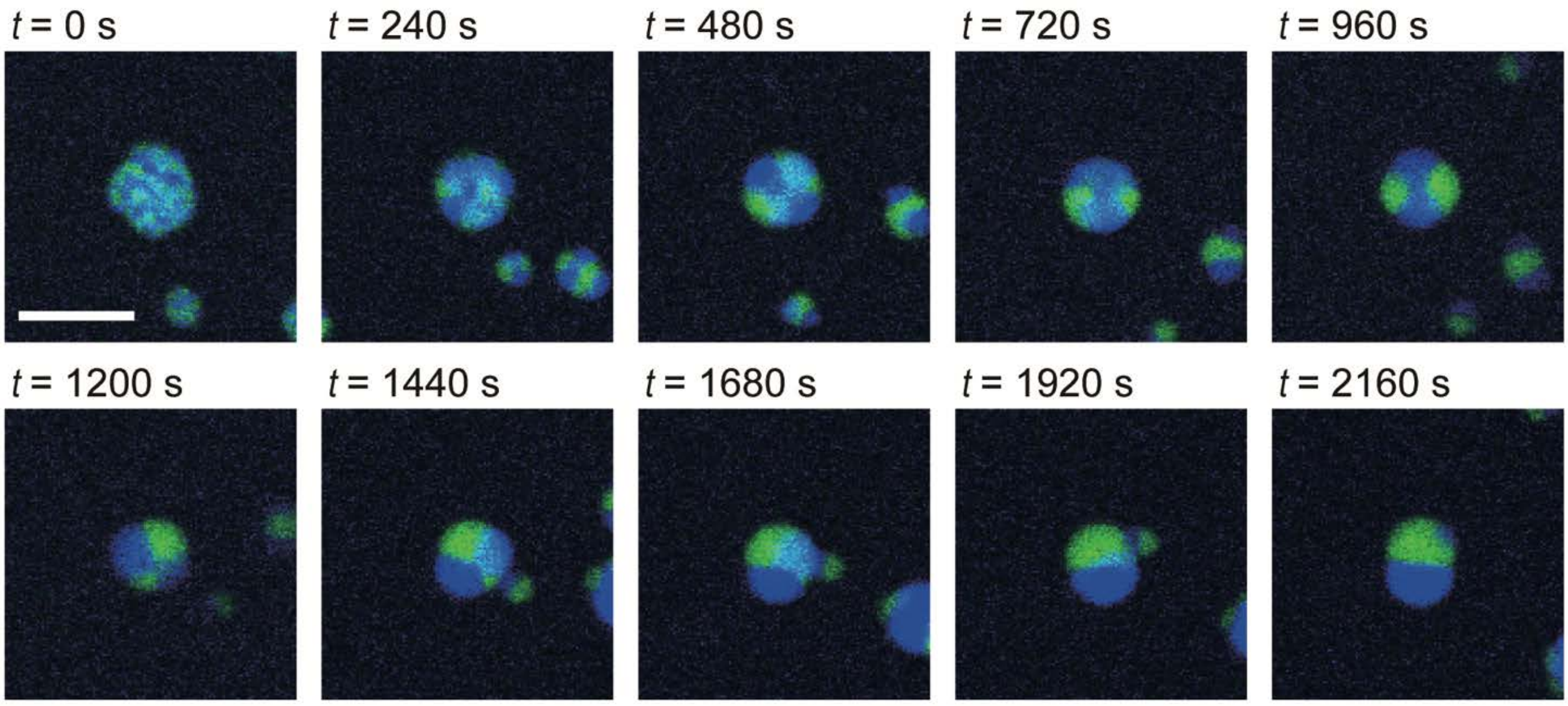

**Supplementary Fig. 25**. Sequential microscopy images of the formation process of the Janus-shape. Blue and green channels indicate ${}^{orth}$Y- (Alexa405) and Y-motifs (FAM), respectively. Scale bar: 20 μm.

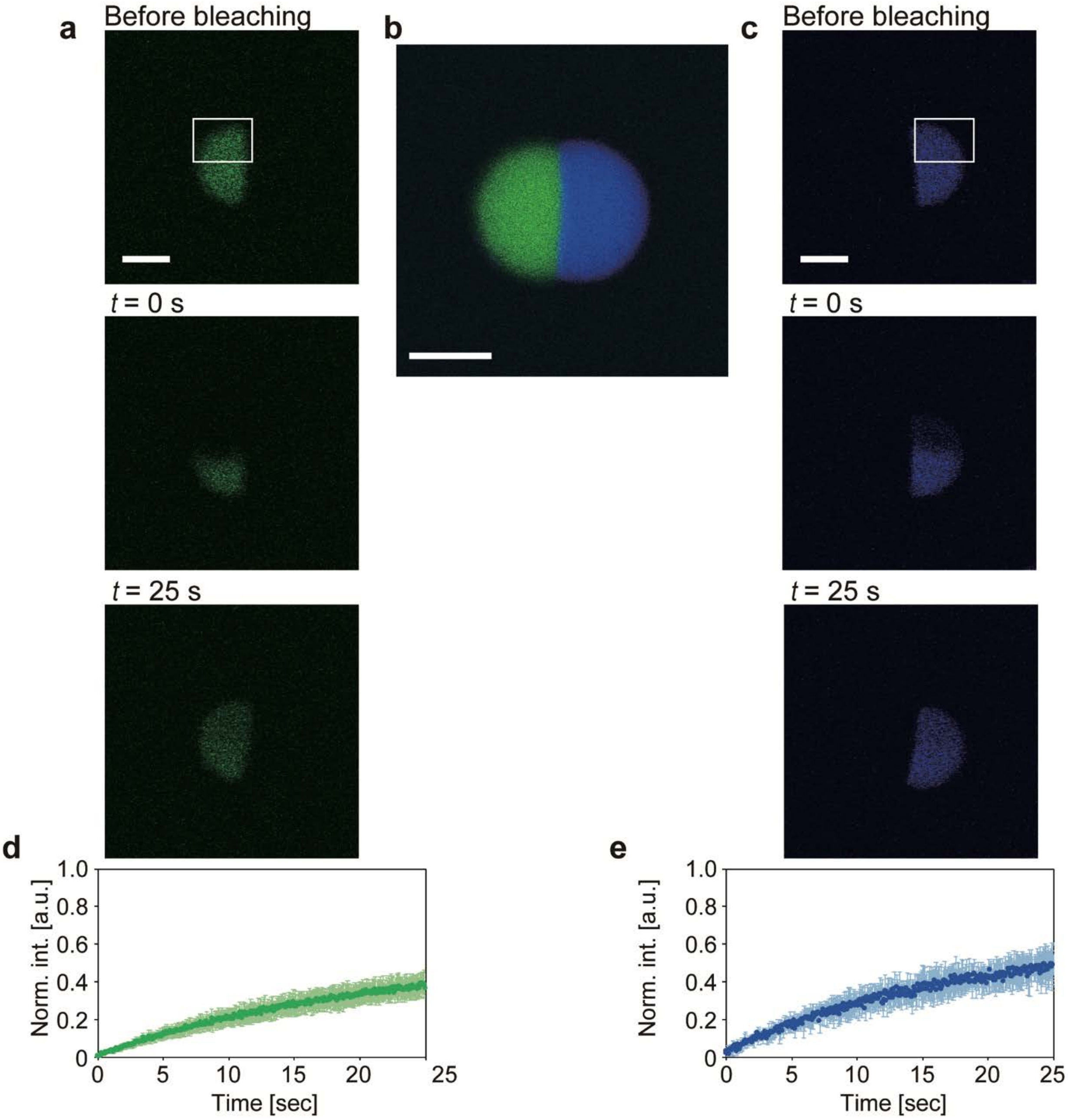

**Supplementary Fig. 26**. FRAP experiments to confirm the fluidity in both motifs in Janus-droplets. **a**, and **c**, Microscopy images from FRAP experiments for FAM-labelled Y-motifs (**a**) and Alexa405-labelled $^{orth}$Y-motifs (**c**). The bleached region is indicated by a white box in each image. **b**, Microscopy images of Janus-droplets before photo-bleaching. Scale bars: 10 μm. **d,** and **e**, Time series of fluorescence intensity of FAM-labelled Y-motifs (**d**) and Alexa405-labelled $^{orth}$Y-motifs (**e**). Error bars show standard deviation ($n$ = 3).

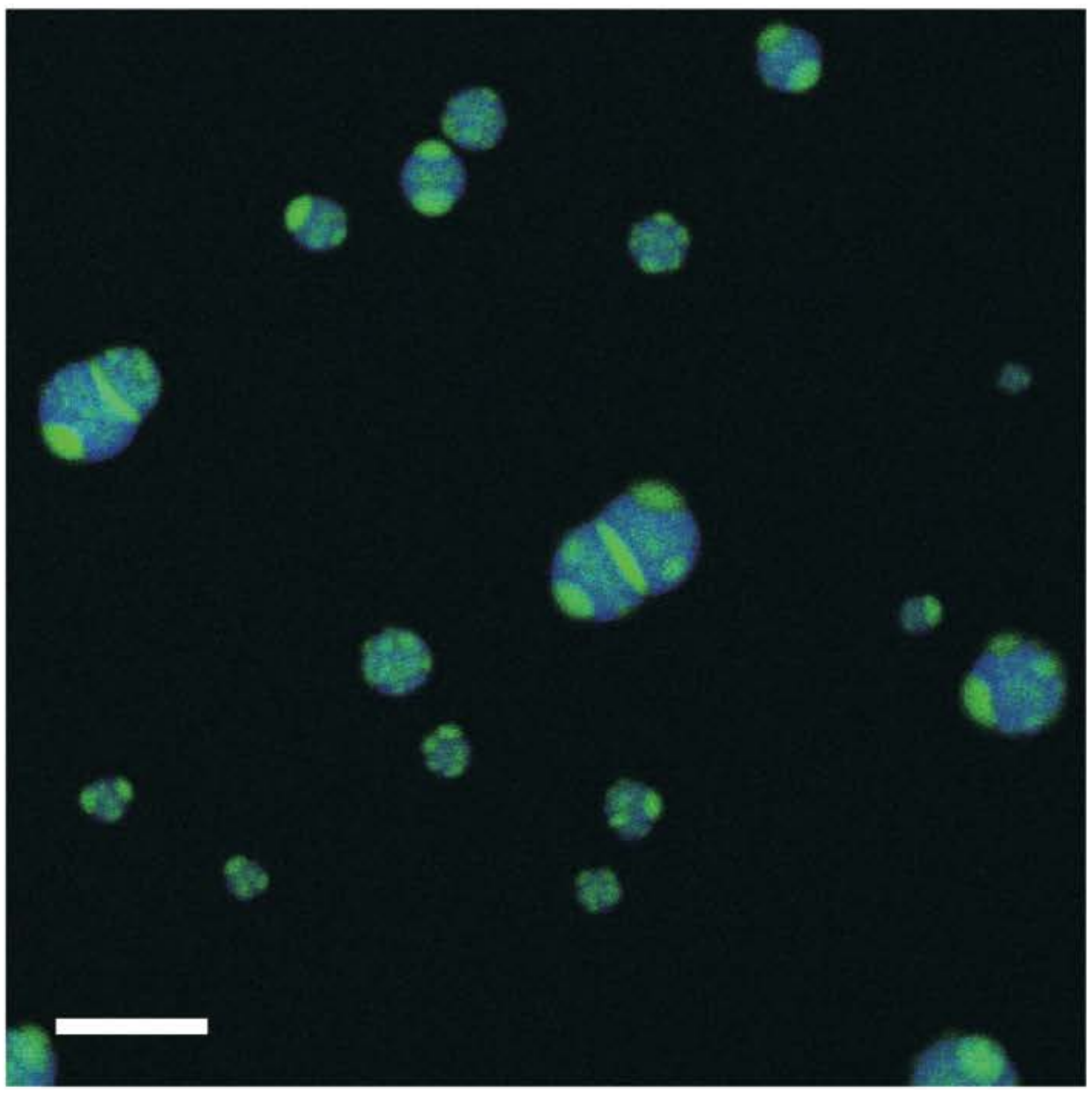

**Supplementary Fig. 27**. Patchy-like patterns in DNA droplets. These were formed with 50 % DNA-RNA chimera strands in the CS-motifs after enzymatic reaction. Blue and green channels indicate $^{orth}$Y- (Alexa405) and Y-motifs (FAM), respectively. The image was obtained 2 h after addition of RNase A. Scale bar: 30 µm.

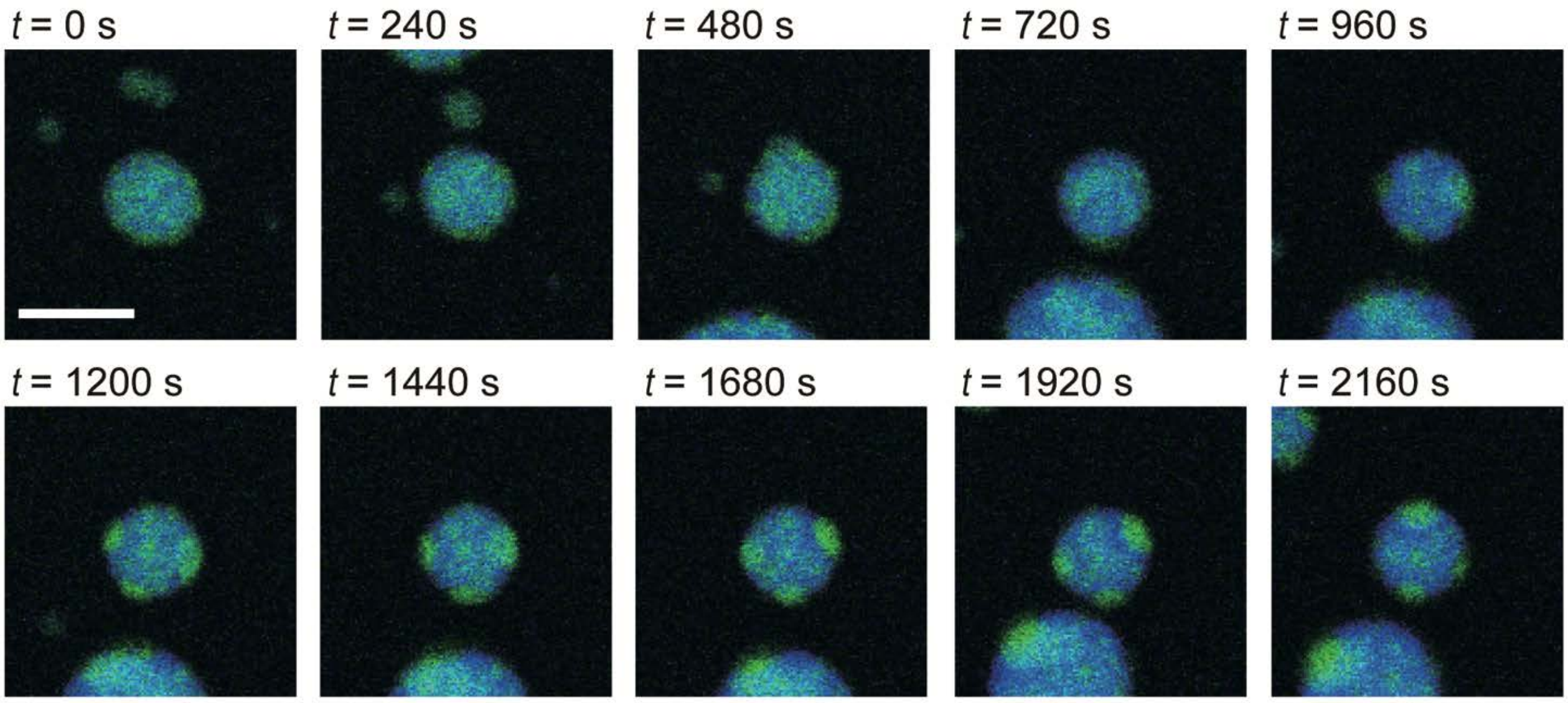

**Supplementary Fig. 28**. Sequential microscopy images of the formation process of patch-like patterns. Blue and green channels indicate $^{orth}$Y- (Alexa405) and Y-motifs (FAM), respectively. Scale bar: 20 µm.

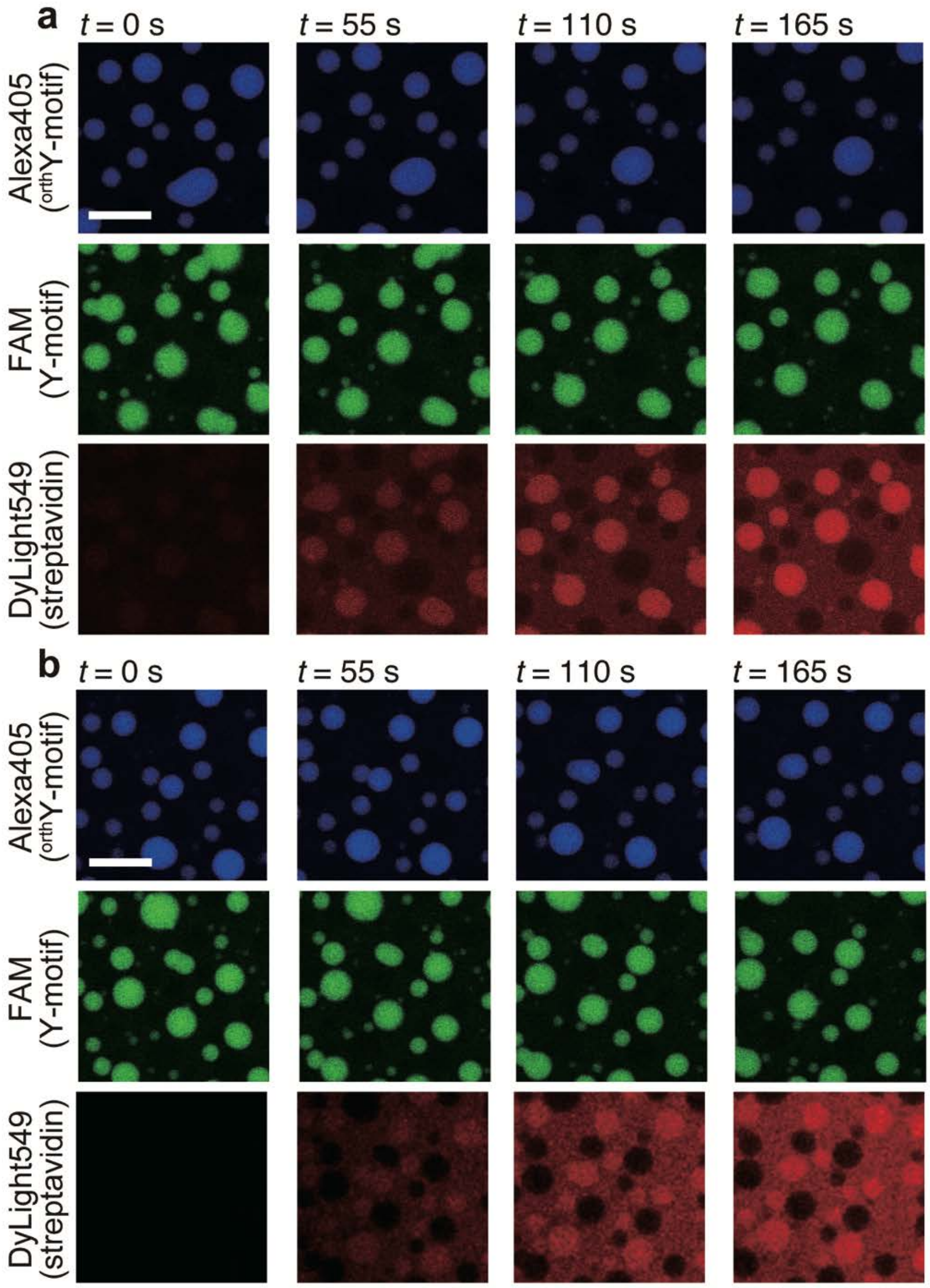

**Supplementary Fig. 29**. Sequential microscopy images of the accumulation process of DNA-modified streptavidin in DNA droplets composed of Y- (**a**) and $^{orth}$Y-motifs (**b**). Scale bars: 50 μm.

## Captions for Supplementary Movies

**Supplementary Movie 1**. Fusion of DNA droplets composed of Y-motifs with 8 nucleotide long sticky ends. This movie was obtained at 63 °C. Scale bar: 30 μm.

**Supplementary Movie 2**. Selective and exclusive fusion of DNA droplets composed of Y- or orthogonal Y-motifs with 8 nucleotide long sticky ends. Sequences of DNAs for both motifs were not complementary. Blue and green channels indicate orthogonal Y-(Alexa405) and Y-motifs (FAM), respectively. This movie was obtained at 64 °C. Scale bar: 30 μm.

**Supplementary Movie 3**. Fission of DNA droplets composed of Y-, orthogonal Y-, and chimerized-six-junction motifs with 8 nucleotide long sticky ends. Blue and green channels indicate orthogonal Y-(Alexa405) and Y-motifs (FAM), respectively. This movie was obtained at 65 °C. Filming began 10 sec after addition of RNase A. Scale bar: 20 μm.